\documentclass[preprint,nofootinbib,aps,superscriptaddress,eqsecnum]{revtex4}
\usepackage{color}
\usepackage{mathtools}
\usepackage{array}
\usepackage{tabularx}
\usepackage{diagbox}
\usepackage{tikz}
\usetikzlibrary{decorations.pathmorphing}
\usepackage{bm}
\usepackage{amsmath,bm,amssymb,amsfonts,dcolumn,color,graphicx,latexsym,epsfig}
\usepackage{rsfso}
\usepackage{hyperref}
\hypersetup{
colorlinks=true,
linkcolor=magenta,
filecolor=magenta,      
urlcolor=blue,
citecolor=blue,
}
\usepackage{subfigure}
\usepackage{multirow}
\usepackage{natbib}
\usepackage{float}
\usepackage{booktabs}
\usepackage{pifont}
\usepackage{mathrsfs}
\usepackage{caption}
\usepackage{gensymb}
\usepackage{caption, threeparttable}
\begin{document}

\title{New rotating Lorentzian wormhole spacetime}
\author{Anjan Kar}
\email[Email address: ]{anjankar.phys@gmail.com}
\affiliation{Department of Physics, Indian Institute of Technology, Kharagpur 721 302, India}
\author{Soumya Jana}
\email[Email address: ]{soumyajana.physics@gmail.com}
\affiliation{Department of Physics, Sitananda College, Nandigram, 721 631, India}
\affiliation{Department of Physics, Indian Institute of Technology, Kharagpur 721 302, India}
\author{Sayan Kar}
\email[Email address: ]{sayan@phy.iitkgp.ac.in}
\affiliation{Department of Physics, Indian Institute of Technology, Kharagpur 721 302, India}
\begin{abstract}
A rotating version of a known static, spherically symmetric, zero Ricci scalar 
Lorentzian wormhole is constructed. It turns out that for this given non-rotating geometry, the standard Newman-Janis 
algorithm
does not produce a rotating wormhole and, therefore, the method 
pioneered by Azreg-A\"inou has to be used. The rotating spacetime thus obtained
is shown to be regular with wormhole features, though it is no longer
a $R=0$ spacetime. The required matter is found to violate the energy conditions,
as expected. A few other characteristic properties of this new rotating spacetime are 
mentioned. Finally, we calculate the shadow for this geometry and 
discuss its features {\em vis-a-vis} the Kerr geometry and available event horizon telescope observations.
\noindent 
\end{abstract}


\maketitle

\newpage

\section{Introduction}
\noindent Lorentzian wormholes have been around for quite a long time now. They are theoretical constructs in theories of gravity and have attracted interest because of various reasons. Among them the prominent ones are (a) their existence versus the violation of energy conditions \cite{Morris1}, (b) the curious possibility of using them to build time machine models \cite{Morris2}, (c) their role as black hole mimickers \cite{damour} especially in the context of gravitational wave observations \cite{Aneesh}, (d) their support for alternative theories of gravity \cite{radhak} and (e) their relevance in issues related to quantum information \cite{kundu}.

\noindent The original wormhole idea goes back to the work of Flamm in 1916 \cite{Flamn} and then Einstein-Rosen in 1935 \cite{Einstein}. The term `wormhole' was coined by Wheeler while discussing `geons' \cite{Wheeler}. A resurgence happened in 1988 through the seminal papers by Morris, Thorne and Morris, Thorne, Yurtsever \cite{Morris1, Morris2}. Subsequently, over the last three and a half decades, wormholery has been largely based on the Morris-Thorne work with later important inputs by various authors (see \cite{Visser, Lobo, Lobo2, Bronnikov1} for literature reviews). 

\noindent Most of the works on wormholes till date have revolved around static, spherically symmetric spacetimes \cite{Morris1}. The first construction of a rotating wormhole was by Teo, in 1998 \cite{Teo}. Since then there have been numerous other constructions too \cite{Kuhfittig, Kashargin, Dzhunushaliev, Bronnikov, Kleihaus, Cisterna, Xio1, Xio3}. Using the rotating wormhole, researchers have carried out studies on various physical and observationally relevant aspects such as lensing, shadows, quasinormal modes, echoes etc. \cite{Shaikh, Jusufi, Deligianni, Xio2, Bueno, Ono, Dejan1, Dejan2, Dejan3, battista1, battista2, abdujabbarov, falco_2023}. The status of energy condition violation for the rotating spacetimes remain largely the same as for their non-rotating counterparts.

\noindent Though briefly mentioned first in \cite{havas}, it was realised in 2002 that imposing only the $R=0$ constraint on a general, static spherically symmetric line element results in a non-vacuum two parameter spacetime different from the vacuum Schwarzschild \cite{nkdsk}. This spacetime, known as a $R=0$ wormhole spacetime was also noted as a solution in a braneworld context by Casadio et al, around the same time \cite{casadio}. Much later, various features of this spacetime have been discussed by different authors --  these include the non-violation of energy conditions for these spacetimes in a braneworld gravity model, quasinormal modes etc \cite{skslssg, Aneesh, IbScSs, Sarkar, RsSk1, RsSk2, IcSbSc}. However, the standard Newman-Janis procedure for finding a rotating version of this
line element does not quite produce a wormhole, as we will see later in this article. 
Hence, a rotating generalisation remained an unresolved issue. 
In this paper, we intend to resolve this issue and construct a rotating
version of this $R=0$ spacetime, which does represent a wormhole.
The geometry of this new spacetime, the matter required as well as
the shadow structure will be explored in detail.

\noindent Our paper is organised as follows. In Section II, we review the
static, spherically symmetric $R=0$ spacetime. Section III first points out
why the standard route of obtaining the rotating version did not work and then
constructs the rotating generalisation using a different method. In Section IV,
we focus on some of the physical properties of the new line element.
Shadows of the new line element and matching with EHT observations are discussed
in Section V. Finally, we present our conclusions in Section VI.

\section{The $R=0$ static wormhole geometry}

\noindent Let us begin by recalling the static, spherically symmetric
spacetime which forms the basis of this work \cite{havas, nkdsk, casadio,skslssg}. 
We begin with the general line element given as,
\begin{eqnarray}
ds^2 = -e^{2\psi (r)} dt^2 + \frac{dr^2}{1-\frac{b(r)}{r}} + r^2 \left (d\theta^2 +\sin^2\theta d\phi^2 \right )
\end{eqnarray}
Evaluating the Ricci scalar $R$ for this spacetime and imposing $R=0$ gives the following
condition on $\xi =\psi'$ (prime denoting a derivative w.r.t. $r$):
\begin{eqnarray}
\xi' +\xi^2 + \left (\frac{2}{r} - \frac{b'r-b}{2r(r-b)} \right ) \xi = \frac{b'}{r(r-b)}
\end{eqnarray}
One cannot however find a solution unless either $\psi (r)$ or $b (r)$ is specified. If we choose $b(r) = 2M$ the R. H. S. is zero and we get the simpler equation
\begin{eqnarray}
\xi' +\xi^2 + \frac{\xi}{r}\left ( \frac{2r-3M}{r-2M}\right ) =0.
\end{eqnarray}
which is easy to solve. The final line element turns out to be
\begin{eqnarray}
    ds^2 = - \left (\kappa +\lambda \sqrt{1-\frac{2M}{r}}\right )^2 dt^2
    +\frac{dr^2}{1-\frac{2M}{r}} +r^2 \left (d\theta^2 +\sin^2\theta d\phi^2\right )
  \end{eqnarray}
the above line element can be recast into a
slightly different form by redefining $\kappa/\lambda= p$ and the scaling $t \rightarrow t/(\lambda (p+1)) $,
\begin{eqnarray}
    ds^2 = -(p+1)^{-2} \left (p +\sqrt{1-\frac{2M}{r}}\right )^2 dt^2
    +\frac{dr^2}{1-\frac{2M}{r}} +r^2 \left (d\theta^2 +\sin^2\theta d\phi^2\right )
    \label{eq:static_wormhole_R0}
\end{eqnarray}

\noindent This is the $R=0$ geometry we wish to work with, in the rest of the article. It is known that the geometry is not a vacuum spacetime and the matter
stress energy, assuming General Relativity, violates the energy conditions \cite{nkdsk}. However, interestingly in a particular braneworld gravity we
find that the same geometry can exist with the `matter' satisfying 
the null energy condition \cite{skslssg}.

\noindent Our central aim in this work is to first find a rotating version of this spacetime. An earlier attempt \cite{sumanta} on the construction of a rotating
generalisation was not quite successful.

\section{The rotating version of the $R=0$ static wormhole}

\subsection{Newman-Janis method and its consequences}
\noindent Beginning with a spherically symmetric, static line element one may obtain a rotating spacetime by applying the Newman-Janis (NJ) algorithm \cite{NJ1,NJ2} systematically. This well-known method was used by the authors of \cite{sumanta} to obtain a rotating spacetime from the static spacetime metric in Eq.~(\ref{eq:static_wormhole_R0}). Below, we briefly describe why the standard NJ algorithm leads to problems, in this specific case.

\noindent The NJ algorithm comprises following steps:\\
$(i)$ For a line element having a generic form:
\begin{equation}
    \text{d}s^2= -f(r)\text{d}t^2+ \frac{\text{d}r^2}{g(r)}+h(r)\left(\text{d}\theta^2+\sin^2\theta \text{d}\phi^2\right),
\end{equation}
one begins by rewriting it using Eddington-Finkelstein coordinates ($u,r,\theta,\phi$), thereby obtaining
\begin{equation}
    \text{d}s^2= -f(r)\text{d}u^2- 2\sqrt{\frac{f}{g}}\text{d}u\text{d}r + h(r)\left(\text{d}\theta^2+\sin^2\theta \text{d}\phi^2\right),
\end{equation}
where $\text{d}t=\text{d}u+ \text{d}r/\sqrt{fg} $.

\noindent $(ii)$ Thereafter, we construct the inverse metric $g^{\mu\nu}$ 
\begin{equation}
    g^{\mu\nu}=- l^{\mu}n^{\nu}-l^{\nu}n^{\mu}+ m^{\mu}\bar{m}^{\nu}+ m^{\nu}\bar{m}^{\mu},
\end{equation}
where $Z^{\mu}_{\alpha}=\left(l^{\mu},n^{\mu},m^{\mu},\bar{m}^{\mu}\right)$ is a set of null tetrad vectors satisfying the relations: $l_{\mu}l^{\mu}=n_{\mu}n^{\mu}=m_
{\mu}m^{\mu}=l_{\mu}m^{\mu}=n_{\mu}m^{\mu}=0$ and $l_{\mu}n^{\mu}=-m_{\mu}\bar{m}^{\mu}=-1$. The tetrad  satisfying these relations is given by
\begin{equation}
    l^{\mu}=\delta^{\mu}_r, \quad~ n^{\mu}=\sqrt{\frac{g}{f}}\delta^{\mu}_{u}-\frac{g}{2}\delta^{\mu}_r, \quad~ m^{\mu}=\frac{1}{\sqrt{2h}}\left(\delta^{\mu}_{\theta}+\frac{i}{\sin \theta}\delta^{\mu}_{\phi}\right),    
\end{equation}
and $\bar{m}^{\mu}$, which is the complex conjugate of $m^{\mu}$.\\

\noindent $(iii)$ The third step is to apply a complex transformation given by 
\begin{equation}
    r'=r+ i a \cos \theta, \quad~ u'= u- i a \cos \theta,
    \label{eq:complex_trans}
\end{equation}

\noindent where $a$ is the rotation parameter. Using the complex transformation, we obtain the new tetrad vectors as, 
\begin{equation}
    l'^{\mu}=\delta^{\mu}_r, \quad~ n'^{\mu}=\sqrt{\frac{G(r,\theta)}{F(r,\theta)}}\delta^{\mu}_{u}-\frac{G(r,\theta)}{2}\delta^{\mu}_r, \quad~ m'^{\mu}=\frac{1}{\sqrt{2H(r,\theta)}}\left(\delta^{\mu}_{\theta}+\frac{i}{\sin \theta}\delta^{\mu}_{\phi} + i a\sin \theta \left(\delta^{\mu}_u-\delta^{\mu}_r\right)\right),    
\end{equation}
where $F(r,\theta)$, $G(r,\theta)$, $H(r,\theta )$ are complexified metric functions. There is no unique way to perform the complexification. However, the most common route employed is to replace the terms $r^2$ by $r'\bar{r'}=\Sigma= r^2+a^2\cos^2\theta$ and $2/r$ by $1/r'+1/\bar{r}'=2r/\Sigma =  2r/(r^2+a^2\cos^2\theta)$. The new inverse metric is thus obtained  using
\begin{equation}
    \tilde{g}^{\mu\nu}=- l'^{\mu}n'^{\nu}-l'^{\nu}n'^{\mu}+ m'^{\mu}\bar{m'}^{\nu}+ m'^{\nu}\bar{m'}^{\mu}.
    \label{eq:inverse_metric}
\end{equation}

\noindent $(iv)$ The fourth and the last step in the algorithm 
is to express the new metric in Boyer-Lindquist coordinates using
a local coordinate transformation defined by
\begin{equation}
    \text{d}u= \text{d}t' + \chi_1(r)\text{d}r, \quad~ \text{d}\phi = \text{d} \phi' +\chi_2(r) \text{d}r   
\end{equation}
such that only $g_{t'\phi'}\neq 0$, and $g_{t'r}=0$, $g_{r\phi'}=0$.
This requirement  fixes 
\begin{equation}
    \chi_1(r)= -\frac{\sqrt{\frac{G(r,\theta)}{F(r,\theta)}}H(r,\theta)+a^2\sin^2\theta}{G(r,\theta)H(r,\theta)+a^2\sin^2\theta}, \quad~ \chi_2(r)= -\frac{a}{G(r,\theta)H(r,\theta)+a^2\sin^2\theta}.
\end{equation}
Further details are available in \cite{rajibul_NJ}. 

\noindent In our case,  $f(r)=(p+1)^{-2}\left(p+\sqrt{1-\frac{2M}{r}}\right)^2$ , $g(r)=1-\frac{2M}{r}$, and $h(r)=r^2$. Thus,  $F(r,\theta)=(p+1)^{-2}\left(p+\sqrt{1-\frac{2Mr}{\Sigma}}\right)^2$, $G(r,\theta)=1-\frac{2Mr}{\Sigma}$, $H(r,\theta)=\Sigma=r^2+a^2\cos^2\theta$. Clearly, unlike the $p=0$ case, the R.H.S. of equation for $\chi_1$ for $p\neq 0$ is not just a function of $r$ but a function of both $r$ and $\theta$. A valid coordinate transformation is therefore non-existent. This was earlier pointed out in \cite{sumanta}.  

\noindent Alternatively, one may first try to use a transformation $r(\ell)$ such that we obtain the transformed metric in the form 
\begin{equation}
    \text{d}s^2= -f(\ell)\text{d}t^2 + \frac{\text{d}\ell^2}{f(\ell)} + h(\ell)\left(\text{d}\theta^2+\sin^2\theta \text{d}\phi^2\right),
    \label{eq:fequalsg}
\end{equation}
In these coordinates $F(\ell,\theta)/G(\ell,\theta)=1$ and 
it may be possible to  have $\chi_1$ and $\chi_2$ as functions of $\ell$ 
alone. In our case, this leads to the transformation equation
\begin{equation}
    \frac{\text{d}\ell}{\text{d}r}= \frac{p+\sqrt{1-\frac{2M}{r}}}{(p+1)\sqrt{1-\frac{2M}{r}}},
    \label{eq:dlbydr}
\end{equation}
which, after solving, gives 
\begin{eqnarray}
  r&=& 2 M \cosh^2 \alpha \\
  \ell &=& \frac{2M}{p+1}\left(\cosh^2\alpha + p \left(\alpha + \sinh \alpha \, \cosh \alpha \right)\right),
\end{eqnarray}
where both $r$ and $\ell$ can be parameterized w.r.t. $\alpha $ but,
unfortunately, the relations are not invertible and we cannot write $r=r(\ell)$ explicitly. Thus, this approach also does not help in producing a desired rotating spacetime which can be written down explicitly using standard coordinates.

\noindent A more general complexification (than the one used in the original NJ algorithm step-$(iii)$) was introduced by Drake and Szekeres \cite{Drake} and the resulting rotating metric is given by \cite{Beltracchi} 
\begin{equation}
\begin{split}
    \text{d}s^2&= -\Sigma \frac{a^2\cos^2\theta +r^2 g}{\Sigma_j^2}\text{d}t^2+ 2\sin^2\theta \frac{(g-j)r^2a\Sigma}{\Sigma_j^2}\text{d}t\text{d}\phi +\frac{\Sigma}{\Delta} \text{d}r^2 +\Sigma \text{d}\theta^2\\
    &+\Sigma \sin^2\theta \frac{(a^2+jr^2)^2-a^2\Delta\sin^2\theta}{\Sigma^2_j}\text{d}\phi^2,
    \end{split}
\end{equation}
where $\Sigma = r^2 + a^2\cos^2\theta$, \, $\Delta= r^2g(r) + a^2$, \, $\Sigma_j= r^2 j(r) + a^2\cos^2\theta $, \, $f(r)=g(r)/[j(r)]^2$, and $h(r)=r^2$. 

\noindent In our case, 
\begin{equation}
f(r)=(p+1)^{-2}\left(p+\sqrt{1-\frac{2M}{r}}\right)^2, \quad~ g(r)= 1- \frac{2M}{r}, \quad~ j(r)=\frac{(p+1)\sqrt{1-2M/r}}{p+\sqrt{1-2M/r}}.
\end{equation}
Although the resulting line element describes a rotating spacetime, this is neither a wormhole nor a black hole. This is evident from $\Delta(r)= r^2+a^2-2Mr$ which has two roots $r_{\pm}=M\pm \sqrt{M^2-a^2}$. Both
the roots $r_{\pm}<2M$ and, therefore, lie 
outside the domain $r\in [2M,\infty)$. Moreover, the curvature scalar for this rotating spacetime does not vanish, i.e. $R\neq0$. Therefore, we do not consider this line element in the remaining parts of this article.

\subsection {Using the Azreg-A\"inou procedure}
\noindent We have shown above how the NJ algorithm 
is unsuccessful in generating a physically meaningful rotating 
wormhole spacetime. In such situations 
one may try to use the Azreg-A\"inou (AA) method \cite{AzregPLB,AzregEPJC,AzregPRD}.  The AA method avoids the ambiguous complexification procedure (step ($iii$) of the  NJ algorithm) and uses a more general approach. The procedure is described briefly below.\\

\noindent Steps $(i)$ and $(ii)$ of the AA method are the same as in the NJ algorithm. The difference occurs in the step $(iii)$ where the complex transformation (Eq.~(\ref{eq:complex_trans})) is applied. Here, it is assumed that $\lbrace f(r), g(r), h(r) \rbrace$ transform to $\lbrace A(r,\theta,a), B(r,\theta, a), \Psi(r,\theta,a) \rbrace$ and the new transformed tetrad with rotation parameter $a$  becomes
\begin{eqnarray}
    l'^{\mu} &=& \delta^{\mu}_r, \label{tetradazreg1}\\
    n'^{\mu} &=& \sqrt{\frac{B}{A}} \delta^{\mu}_u - \frac{B}{2}\delta^{\mu}_r, \label{tetradazreg2}\\
    m'^{\mu} &=& \left. \left(\delta^{\mu}_\theta + i a\sin \theta \, \left(\delta^{\mu}_u- \delta^{\mu}_r\right) + \frac{i}{\sin \theta}\delta^{\mu}_\phi\right) \right/ \sqrt{2\Psi}
    \label{tetrad_azreg}
\end{eqnarray}
where $\lbrace A, B, \Psi \rbrace$ are three-variable real functions to be fixed later.  In the limit $a\rightarrow 0$ these functions approach the correct limits, i.e.
\begin{equation}
    \lim_{a \to 0} A(r,\theta,a)= f(r), \quad~ \lim_{a \to 0} B(r,\theta,a)= g(r), \quad~ \lim_{a \to 0} \Psi(r,\theta,a)= h(r).
\end{equation}
Hence the inverse metric for the new rotating spacetime is given by the Eq.~(\ref{eq:inverse_metric}) where $\lbrace l'^{\mu}, n'^{\mu}, m'^{\mu} \rbrace$ are given by Eqs.~(\ref{tetradazreg1},\ref{tetradazreg2},\ref{tetrad_azreg}). The line element of the new rotating metric in Eddington-Finkelstein coordinates is 
\begin{equation}
\begin{split}
    \text{d}s^2 & = -A \text{d}u^2 -2\sqrt{\frac{A}{B}}\text{d}u\text{d}r + 2\left(A-\sqrt{\frac{A}{B}}\right) a\sin^2 \theta \text{d}u\text{d}\phi + 2\sqrt{\frac{A}{B}} a\sin^2 \theta \text{d}r \text{d}\phi\\
  &  +\Psi \text{d}\theta^2 + \sin^2\theta \left[ \Psi + \left(2\sqrt{\frac{A}{B}}-A\right)a^2\sin^2\theta\right]\text{d}\phi^2.
    \end{split}
\end{equation}
In order to get the rotating metric in Boyer-Lindquist (BL) coordinates, we give the following transformations \cite{AzregEPJC} 
\begin{eqnarray}
    \text{d} u= \text{d} t - \frac{K + a^2}{gh + a^2} \text{d} r , \label{eq:BL_u}\\
    \text{d} \phi = \text{d} \varphi - \frac{a}{g h + a^2}\text{d} r, \label{eq:BL_phi}
\end{eqnarray}
where $K(r)= h\sqrt{g/f} $. Choosing the metric component $g_{tr}=0$ in BL coordinates we get
\begin{equation}
    A \left( K + a^2 \cos^2 \theta \right) -\sqrt{ \frac{A}{B}}\left( g h + a^2\cos^2 \theta \right)= 0.
    \label{eq:g_tr=0}
\end{equation}
and, further, choosing $g_{r\phi}=0$ we get
\begin{equation}
   -  A \left( K + a^2 \cos^2 \theta \right) + \sqrt{\frac{A}{B}} \left( K + g h + 2 a^2\cos^2 \theta  \right) = \Psi .
   \label{eq:g_rphi=0}
\end{equation}
Solving Eqs.~(\ref{eq:g_tr=0}) and (\ref{eq:g_rphi=0}) for $A$ and $B$, we get 
\begin{eqnarray}
    A &=& \frac{g h + a^2 \cos^2 \theta}{ \left(K + a^2 \cos^2 \theta \right)^2} \Psi, \label{eq:A}\\
    B &=& \frac{g h + a^2 \cos^2 \theta}{\Psi}. \label{eq:B}
\end{eqnarray}
Finally, using the Eqs.~(\ref{eq:BL_u}), (\ref{eq:BL_phi}), (\ref{eq:A}), and (\ref{eq:B}), we get the the rotating spacetime in BL coordinates with
its line element given as,
\begin{equation}
\begin{split}
    \text{d} s^2 = & - \frac{g h + a^2 \cos^2 \theta }{\left(K + a^2\cos^2 \theta\right)^2} \Psi \text{d}t^2 + \frac{\Psi}{g h + a^2} \text{d} r^2 -2 a \sin^2 \theta \Psi \left[\frac{K- g h}{\left(K + a^2\cos^2 \theta\right)^2} \right] \text{d} t \text{d} \varphi \\
    & + \Psi \text{d} \theta^2 + \Psi \sin^2 \theta \left[1 + a^2 \sin^2 \theta \frac{\left(2K - g h + a^2\cos^2\theta \right)}{\left(K + a^2\cos^2 \theta\right)^2} \right] \text{d} \varphi^2.
    \end{split}
    \label{eq:rotating_metric_Azreg}
\end{equation}
 We still need to fix the function $\Psi(r,\theta)$. We assume an additional constraint on the rotating axially symmetric spacetime such that the $ r\theta $-component of the Einstein tensor becomes zero, i.e. $G_{r \theta }=0$. This leads to the differential equation for $\Psi$ given as 
\begin{equation}
 \left(K + a^2 \cos^2 \theta\right)^2 \left(3 \Psi_{,r} \Psi_{,\theta} - 2 \Psi \Psi_{,r\theta}\right) + 3a^2 \sin(2\theta) K_{,r}\Psi^2 = 0.
 \label{eq;Psi}
\end{equation}
An immediate solution for $\Psi$ (without any further assumption, for example, on the imperfect fluid as the source of the rotating spacetime \cite{AzregEPJC}) is given as
\begin{equation}
    \Psi(r,\theta)= K + a^2\cos^2\theta .
    \label{eq:Psi_soln}
\end{equation}

\subsection{The new geometry}

\noindent In this section, we apply the AA method to the static line element given by Eq.~(\ref{eq:static_wormhole_R0}) and obtain a new rotating spacetime geometry. We first write the line element in the form given by Eq.~(\ref{eq:fequalsg}) such that $K(\ell)=h(\ell)=r^2(\ell)$ and $d\ell/dr$ is given by Eq.~(\ref{eq:dlbydr}). However, we no longer require an invertible relation between $\ell$ and $r$, which was necessary for applying the NJA method. Then, using the Eqs.~(\ref{eq:rotating_metric_Azreg}) and (\ref{eq:Psi_soln}), we get a new rotating metric
\begin{equation}
\begin{split}
    \text{d}s^2& = -\left[1-\frac{(1-f(\ell))}{\Sigma}r^2(\ell)\right]\text{d}t^2-2\frac{a\sin^2\theta}{\Sigma}(1-f(\ell))r^2(\ell)\text{d}t\text{d}\varphi + \frac{\Sigma}{f(\ell)r^2(\ell)+a^2}\text{d}\ell^2 \\
    & + \Sigma \text{d}\theta^2 + \sin^2\theta \left[r^2(\ell)+a^2 + \frac{a^2\sin^2\theta}{\Sigma}\left(1-f(\ell)\right)r^2(\ell)\right]\text{d}\varphi^2
    \end{split},
\end{equation}
where 
\begin{eqnarray}
    \Sigma &=& r^2(\ell)+a^2\cos^2\theta, \\
    f(\ell) &=& \left(p+1\right)^{-2} \left(p+ \sqrt{1-\frac{2M}{r(\ell)}}\right)^2.
\end{eqnarray}
Finally, we go back from $\ell$ coordinate to $r$ coordinate by using the Eq.~(\ref{eq:dlbydr}) and obtain the rotating metric in BL coordinates:
\begin{equation}
\begin{split}
    \text{d}s^2 &= -\left[1- \frac{2m(r) r}{\Sigma}\right]\text{d}t^2- \frac{4m(r) r a\sin^2\theta}{\Sigma} \text{d}t\text{d}\varphi + \frac{\Sigma}{\Delta} \text{d}r^2 + \Sigma \text{d}\theta^2 \\
    &+ \sin^2\theta \left[ r^2 + a^2 + \frac{2m(r) r a^2\sin^2\theta}{\Sigma} \right]\text{d}\varphi^2,
    \end{split}
    \label{eq:rotating_wormhole_new}
\end{equation}
where
\begin{eqnarray}
    m(r) &=& \frac{r}{2} \left[1-\frac{\left(p+ \sqrt{1-2M/r}\right)^2}{(p+1)^2}\right],\label{eq:m(r)}\\
    \Delta(r) &=& \left(1-\frac{2M}{r}\right)\left[\frac{\left(p+ \sqrt{1-2M/r}\right)^2r^2 + a^2(p+1)^2}{\left(p+ \sqrt{1-2M/r}\right)^2}\right], \label{eq:Delta(r)}
\end{eqnarray}
and $\Sigma= r^2 + a^2\cos^2\theta  $.

\section{Physical properties of the new rotating spacetime}

\subsection{Geometry, wormhole character}
\noindent In the static limit $a=0$, we get back the $R=0$ static wormhole [Eq.~(\ref{eq:static_wormhole_R0})] from the line element Eq.~(\ref{eq:rotating_wormhole_new}). Also, in the $p=0$ limit, we get back the Kerr black hole. The resulting rotating spacetime describes a rotating wormhole. The metric (\ref{eq:rotating_wormhole_new}) can be recast in the generic form of the rotating wormhole spacetime as
\begin{equation}
    \text{d}s^2= - N^2\text{d}t^2 + \frac{\text{d}r^2}{1- b/r} + r^2 \tilde{K}^2\text{d}\theta^2 + r^2K^2\sin^2 \theta \left(\text{d}\varphi - \omega \text{d}t\right)^2 ,
    \label{eq:rotatingWH_generic}
\end{equation}
where $N^2, b, K^2, \tilde{K}^2, \omega$ are all functions of $r$ and $\theta$ only, such that it is regular on the symmetry axis $\theta =0, \pi$. $N^2(r,\theta)$ is analogous to the redshift function and $N^2(r,\theta)> 0$ for all ($r,\theta$) for the non-existence of an event horizon and a curvature singularity. $b(r,\theta)$ is the shape function and, at the wormhole throat $r=r_0= b(r_0,\theta)$, where two identical, asymptotically flat regions are joined together. Additionally, the shape function $b(r,\theta)$ satisfies the conditions: $(i)$ $b(r,\theta)\leq r$, $(ii)$ at the throat $r=r_0$, $b(r,\theta)$ is independent of $\theta$, i.e. $b_\theta=0$ ($b_{\theta}= \frac{\partial b}{\partial \theta}$), and $(iii)$ the flare-out condition, i.e. at $r=r_0$,  $b_r<0$ ($b_r=\frac{\partial b}{\partial r}$). Next, $\omega(r,\theta)$ is the angular velocity $d\varphi/dt$ acquired by a particle that falls freely from infinity to the point $(r,\theta)$. At any point $(r,\theta)$ the `proper radial distance' is given by $rK(r,\theta)$ where $K^2(r,\theta)>0$. Lastly, $\tilde{K}^2(r,\theta)$ is an arbitrary positive definite function. In general $K^2\neq \tilde{K}^2$, but for Teo type wormholes \cite{Teo} $K^2=\tilde{K}^2$.  In our case, the explicit forms of the metric functions are
\begin{eqnarray}
    N^2(r,\theta) &=& \frac{\Sigma \left(r^2 + a^2 -2 m(r)r\right)}{\Sigma\left(r^2 + a^2\right) + 2 m(r) r a^2\sin^2 \theta}, \\
    b(r,\theta) &=& r - \frac{r\Delta(r)}{r^2+ a^2\cos^2 \theta}, \\
    \omega(r,\theta) &=& \frac{2m(r)r a}{ \Sigma\left(r^2 + a^2\right) + 2 m(r) r a^2\sin^2 \theta}, \\
    K^2(r,\theta) &=& 1 + \frac{a^2}{r^2} + \frac{2m(r) a^2\sin^2\theta}{\Sigma r}, \\
    \tilde{K}^2(r,\theta) &=& 1 + \frac{a^2\cos^2\theta}{r^2},
\end{eqnarray}
where $m(r)$, $\Sigma$, and $\Delta(r)$ are, as given in the previous section. One can easily verify that the metric functions satisfy all of the criteria for a wormhole geometry, as stated above. However, our wormhole metric is not of the type found by Teo \cite{Teo}. Rather, it can be classified as the Kerr-like wormholes similar to those proposed in \cite{KerrlikeWH} as the rotating generalization of the Damour-Solodukhin wormhole \cite{DSWH}. 

\noindent Now we can construct the embedding diagrams of our rotating wormhole by considering an equatorial slice $\theta=\pi/2$ and at a fixed $t$. Then the metric takes the form
\begin{equation}
    ds^2 = \frac{dr^2}{1-b(r)/r} + { S}^2(r) d\varphi^2,
    \label{eq:WH_embeddingmetric}
\end{equation}
where 
\begin{equation}
    {S}^2(r)= r^2 + 2 a^2 - \frac{a^2\left(p+ \sqrt{1-2M/r}\right)^2}{(p+1)^2},\quad~ b(r)= 2M - \frac{a^2(p+1)^2(r-2M)}{r^2\left(p+ \sqrt{1-2M/r}\right)^2}.
    \label{eq:shapeWH}
\end{equation}
We embed the metric (\ref{eq:WH_embeddingmetric}) into three-dimensional Euclidean space to visualize this slice. Using the distance in cylindrical coordinates and $S$, $z$ as functions of $r$, we get
\begin{equation}
    ds^2=dz^2 +d {S}^2 + { S}^2 d\varphi^2= \left[ \left(\frac{dz}{dr}\right)^2+  \left(\frac{d {S}}{dr}\right)^2 \right] dr^2 + {S}^2 d\varphi^2.
    \label{eq:WH_embed_cylindrical}
\end{equation}

\begin{figure}[!htbp]
\centering
\subfigure[\hspace{0.1cm}$a/M=0.1,\, p=0.1$]{\includegraphics[width=3.0in]{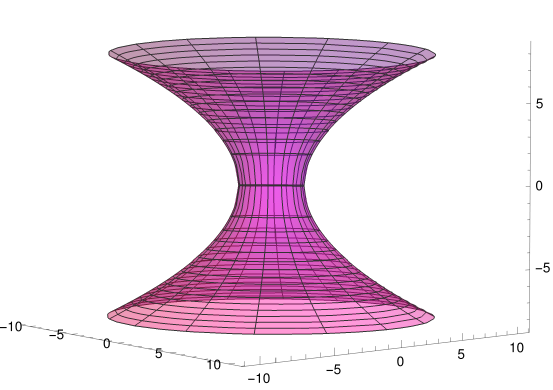}\label{subfig:WHembed_a0.1}}
\subfigure[\hspace{0.1cm}$a/M=0.9,\, p=0.1$]{\includegraphics[width=3.0in]{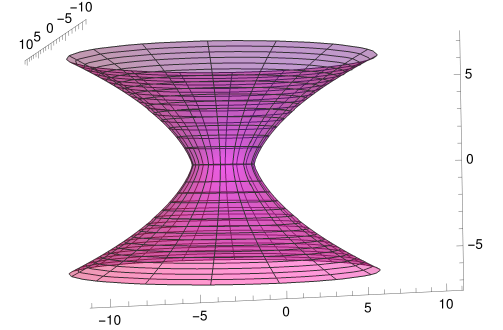}\label{subfig:WHembed_a0.9}}
\caption{Embedding diagram of the rotating wormhole.}
\label{fig:WH embedding}
\end{figure}

\noindent The combination of the Eqs.~(\ref{eq:WH_embeddingmetric}) and (\ref{eq:WH_embed_cylindrical}) gives the equation for the embedded surface 
given as
\begin{equation}
    \frac{dz}{dr} = \pm \sqrt{ \frac{r}{r-b(r)}- \left(\frac{d {S}}{dr}\right)^2},
\end{equation}
where $b(r)$ and $d{S}/dr$ are given by Eq.~(\ref{eq:shapeWH}). This equation cannot be solved analytically. Figure~\ref{fig:WH embedding} shows the embedding diagram of the rotating wormhole after solving the embedding equation numerically. 

\noindent The throat radius of the rotating wormhole is $S\vert_{r=2M}= \left[4M^2 + \frac{a^2(p^2+4p+2)}{(p+1)^2} \right]^{1/2}$, which depends  on both the rotation parameter $a$ and the parameter $p$, in addition to the ADM mass $M$.

\noindent However, one can verify that the curvature scalar of the resulting wormhole spacetime is nonvanishing ($R\neq 0$), which is explicitly shown in the next section. This undesirable feature makes it
somewhat different from the static $R=0$ wormhole geometry with which we
began our journey. 

\noindent A possible generalization of the rotating wormhole metric obtained here, with the aim  of achieving $R=0$, could be
\begin{equation}
\begin{split}
    \text{d}s^2 &= -\left[1- \frac{2m(r) r}{\Sigma}\right]\text{d}t^2- \frac{4m(r) r a\sin^2\theta}{\Sigma} \text{d}t\text{d}\varphi + \frac{\Sigma}{\Delta} \text{d}r^2 + \Sigma \text{d}\theta^2 \\
    &+ \sin^2\theta \left[ r^2 + a^2 + \frac{2m(r) r a^2\sin^2\theta}{\Sigma} \right]\text{d}\varphi^2,
    \end{split}
    \label{eq:rotating_wormhole_ansatz}
\end{equation}
where
\begin{equation}
    \Delta(r)=\left(1-\frac{2M}{r}\right) \left(\frac{r^2-2r m(r) +a^2}{1-2m(r)/r}\right),
    \label{eq:Delta_ansatz}
\end{equation}
and $\left(1-\frac{2m(r)}{r}\right)>0$ for a wormhole geometry.
However, one can check that the Kerr black hole is the unique $R=0$ rotating solution, i.e. $(m(r)=M)$ for such a metric ansatz. 

\noindent In the limit $M\rightarrow 0$, similar to Kerr black holes,  our rotating wormhole metric [Eq.~(\ref{eq:rotating_wormhole_new})] reduces to 
\begin{equation}
    ds^2= -dt^2 + \frac{r^2+a^2\cos^2\theta}{r^2+a^2} dr^2 + \left(r^2 + a^2\cos^2\theta \right) d\theta^2 + (r^2+a^2)\sin^2\theta d\phi^2,
\end{equation}
which is the {\em locally} flat metric in spheroidal coordinates \cite{Gibbons zero mass limit}:
\begin{eqnarray}
    x &=& \sqrt{r^2+a^2}\sin \theta \cos \phi, \nonumber\\
   y &=& \sqrt{r^2+a^2}\sin \theta \sin \phi, \nonumber\\
   z &=& r \cos \theta.
   \label{eq:spheroidal_coord}
\end{eqnarray}



\noindent Another geometrical property worth noting is that 
an ergosphere does not exist for our rotating wormhole. For existence of ergosphere in a rotating spacetime, the metric component $g_{tt}=0$. In our case,
\begin{equation}
    g_{tt}=-1 + \frac{2m(r)r}{r^2+a^2\cos^2\theta}= -\left(\frac{a^2\cos^2\theta + r^2(p+\sqrt{1-2M/r})^2(p+1)^{-2}}{r^2+a^2\cos^2\theta}\right) <0,\quad~ \forall\, r,\theta .
\end{equation}

\subsection{Required matter and energy conditions}
\noindent In the framework of Einstein's GR, the matter required to sustain a geometry must obey the energy conditions, namely null, weak, strong and dominant energy conditions \cite{Hawking, Wald}. Let us now examine these energy conditions for the matter that is required to support our rotating wormhole. As our metric contains an off diagonal term, one cannot 
directly use the energy condition inequalities used for a 
purely diagonal stress-energy. One way to deal with this issue is to choose the observer in a Locally Non-rotating Frame (LNRF) \cite{bardeen_LNRF,bambi_LNRF} for which the basis vectors are:
\begin{equation}
\textbf{e}^{(0)}= \left\vert- g_{tt} + \frac{g_{t\phi}^2}{g_{\phi\phi}} \right\vert^{1/2}\textbf{dt}, \quad~ \textbf{e}^{(1)}= \sqrt{g_{rr}}\,\textbf{dr}, \quad~ \textbf{e}^{(2)}=\sqrt{g_{\theta\theta}}\textbf{d}\,\theta, 
\quad~ \textbf{e}^{(3)}= \frac{g_{t\phi}}{\sqrt{g_{\phi\phi}}}\,\textbf{dt} + \sqrt{g_{\phi\phi}}\,\textbf{d}\phi.
\label{NLRF basis}
\end{equation}

\noindent In the orthonormal basis, the energy-momentum tensor $T^{\alpha\beta}$ admits the decomposition \cite{eric_poisson}
\begin{equation}
    T^{\alpha\beta}= \rho\, \hat{e}^{\alpha}{}_0\hat{e}^{\beta}{}_0 +p_1\, \hat{e}^{\alpha}{}_1\hat{e}^{\beta}{}_1 + p_2\, \hat{e}^{\alpha}{}_2\hat{e}^{\beta}{}_2 + p_3\,  \hat{e}^{\alpha}{}_3\hat{e}^{\beta}{}_3
\end{equation}
where $\rho$ is the energy density and $p_i$ ($i=\lbrace 1,2, 3 \rbrace$) are the principal pressures. $\hat{e}^{\alpha}{}_a$ are the inverse tetrad components such that
$\textbf{e}^{(a)}=\hat{e}_{\mu}{}^a dx^{\mu}$ (here the Latin index $a$ denotes the local inertial frame and have the values $a=\lbrace 0, i \rbrace$). The expressions for $\rho$ and $p_i$ are given as,
\begin{equation}
    \rho = T^{\alpha\beta} \hat{e}_{\alpha}{}^0\hat{e}_{\beta}{}^0, \quad~ p_i= T^{\alpha\beta} \hat{e}_{\alpha}{}^i\hat{e}_{\beta}{}^i
    \label{energy-pressure}
\end{equation}
using the orthogonality relation $\hat{e}_{\mu}{}^a\hat{e}^{\mu}{}_b=\delta^a_b$.

\noindent Then, in the LNRF frame, the energy conditions can be written in the following way,
\begin{equation}
    \begin{aligned}
       \text{Null energy condition:}& \hspace{0.2in} \rho+p_{i}\geq 0 \hspace{0.3in}(i=1,2,3)\\ 
       \text{Weak energy condition:}& \hspace{0.2in} \rho \geq 0, \hspace{0.2in} \rho+ p_{i}\geq 0 \hspace{0.3in}(i=1,2,3)
    \end{aligned}
\end{equation}

\noindent For our case, we use Eqs.(\ref{energy-pressure}) and (\ref{NLRF basis}) for the required $T_{\mu\nu}$ of the rotating wormhole solution. 
Next, we have graphically analysed the null and weak energy conditions at $\theta=\pi/2$.
\begin{figure}[h]
    \centering
    \includegraphics[width=0.326\textwidth]{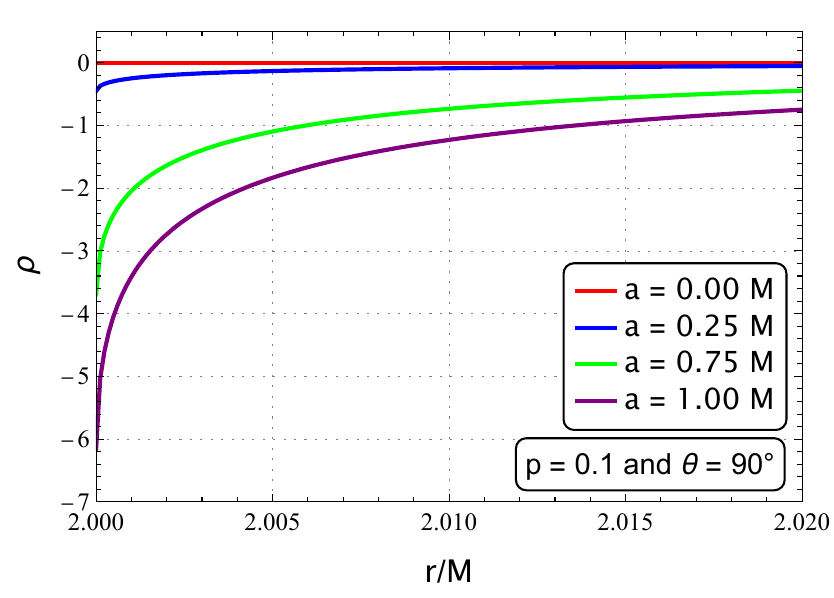}
         \includegraphics[width=0.326\textwidth]{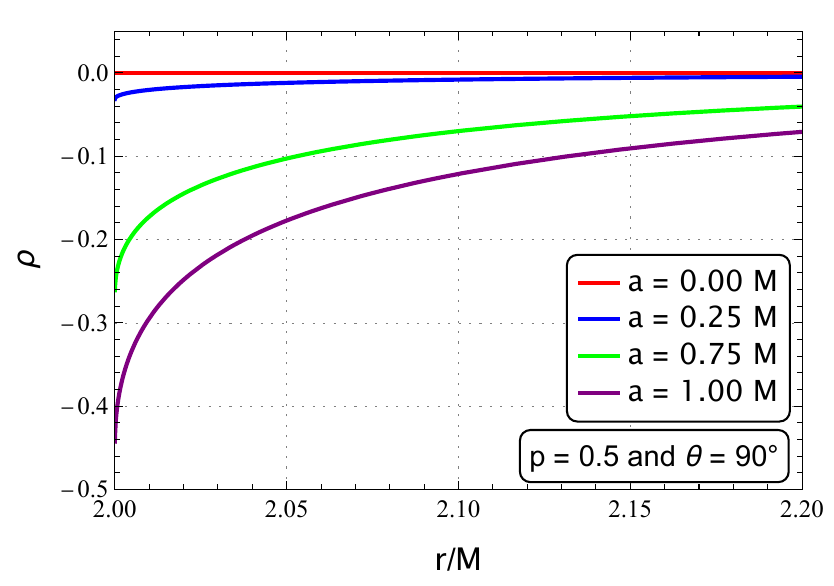}
         \includegraphics[width=0.326\textwidth]{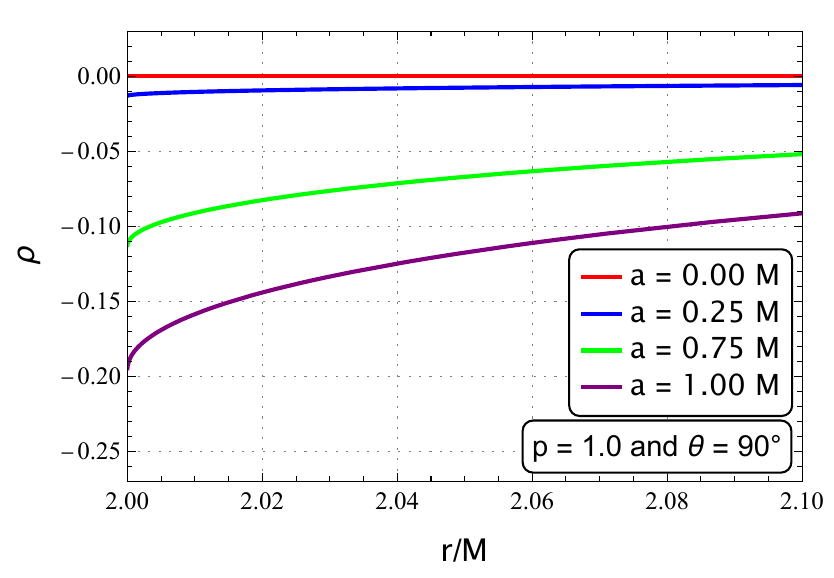}
    \caption{The variation of energy density ($\rho$) is shown w.r.t. $r$ for different values of $a$ and $p$ at $\theta=\pi/2$. }
    \label{fig:WEC}
\end{figure}
In Figure \ref{fig:WEC}, we have shown that the energy density is negative all over $r$ for different values of $a$ and $p$, which confirms the violation of the Weak Energy Condition. Note that $a=0$ represents the original $R=0$ static metric of zero energy density, which is also evident from Figure \ref{fig:WEC}. Moreover, the finite energy density at the throat radius ($r=2M$) for different values of $a$ and $p$ indicates the non-singular character of our geometry. To check the validity of the Null Energy Condition, we have plotted the different 
quantities which arise in the Null Energy Condition inequalities as
a function of the radial coordinate, for various values of $a$ and $p$ at $\theta=\pi/2$, in Figure \ref{fig:NEC}. 
\begin{figure}[h]
         \includegraphics[width=0.326\textwidth]{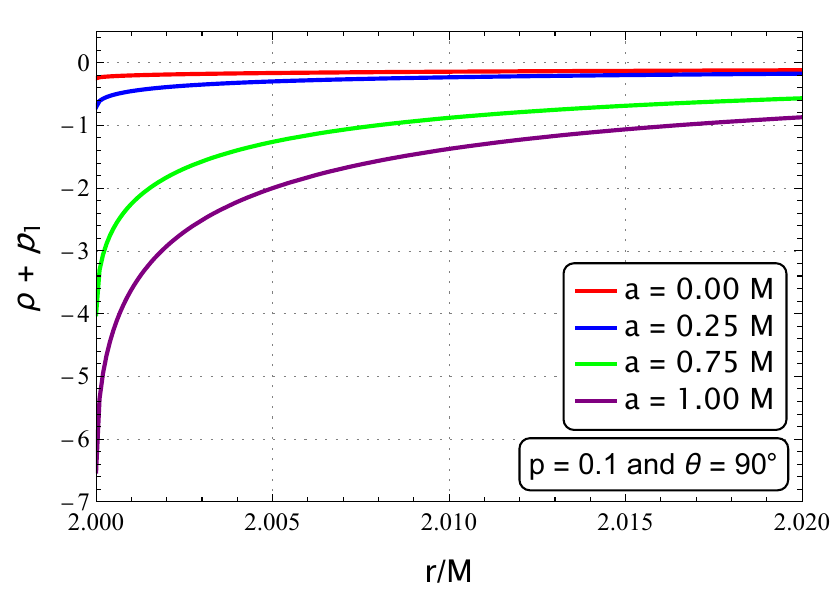}
         \includegraphics[width=0.326\textwidth]{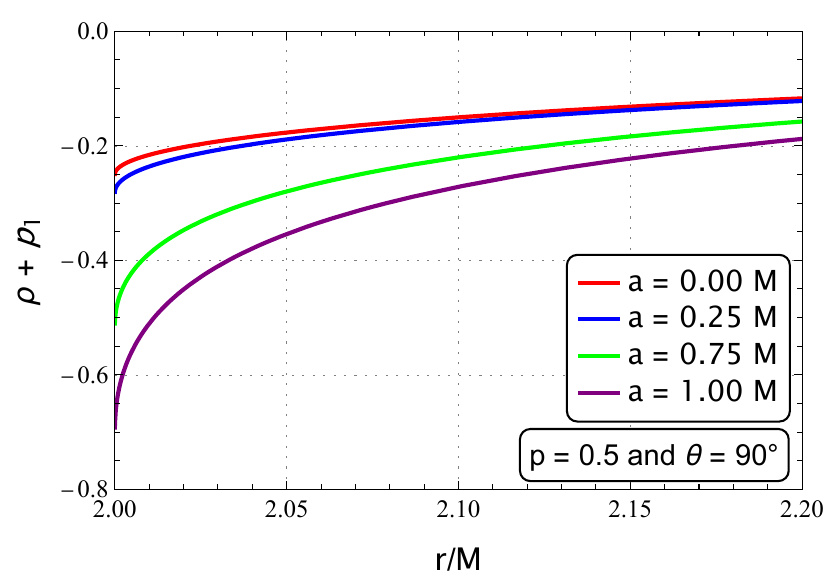}
         \includegraphics[width=0.326\textwidth]{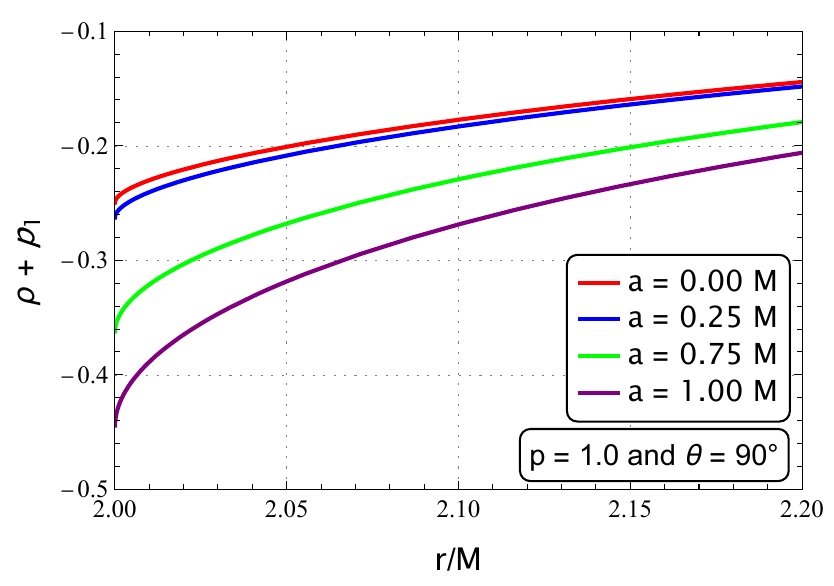}
         \includegraphics[width=0.326\textwidth]{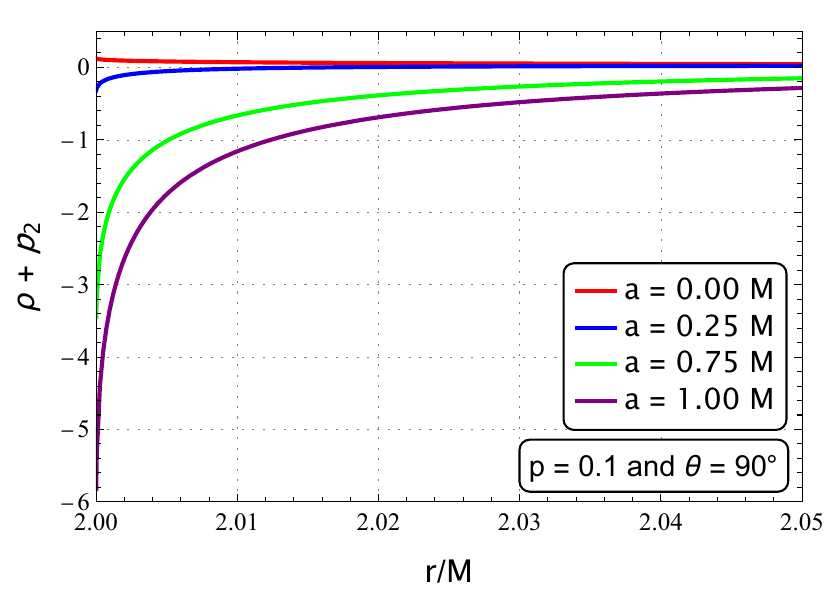}
         \includegraphics[width=0.326\textwidth]{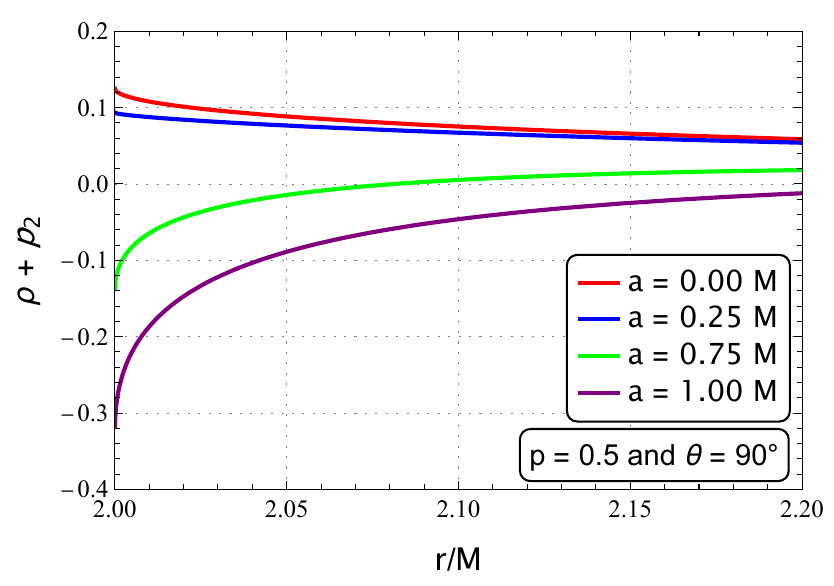}
         \includegraphics[width=0.326\textwidth]{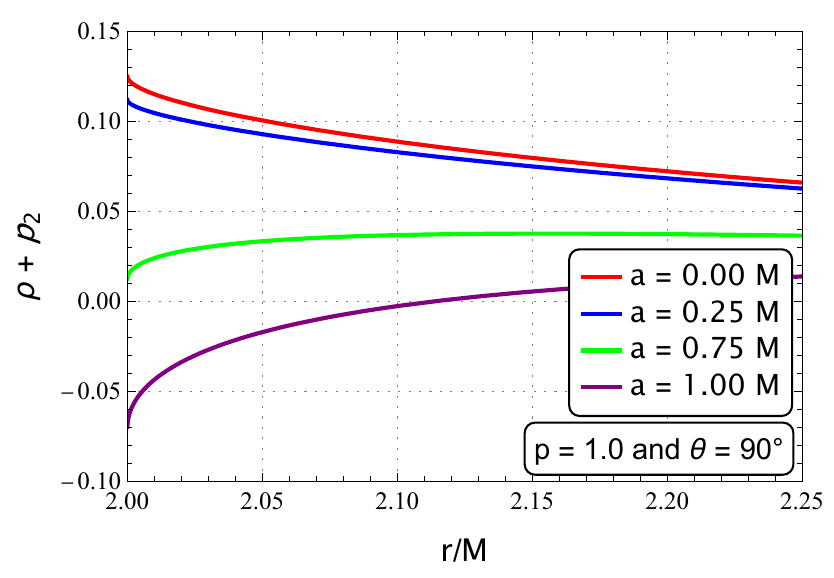}
         \includegraphics[width=0.326\textwidth]{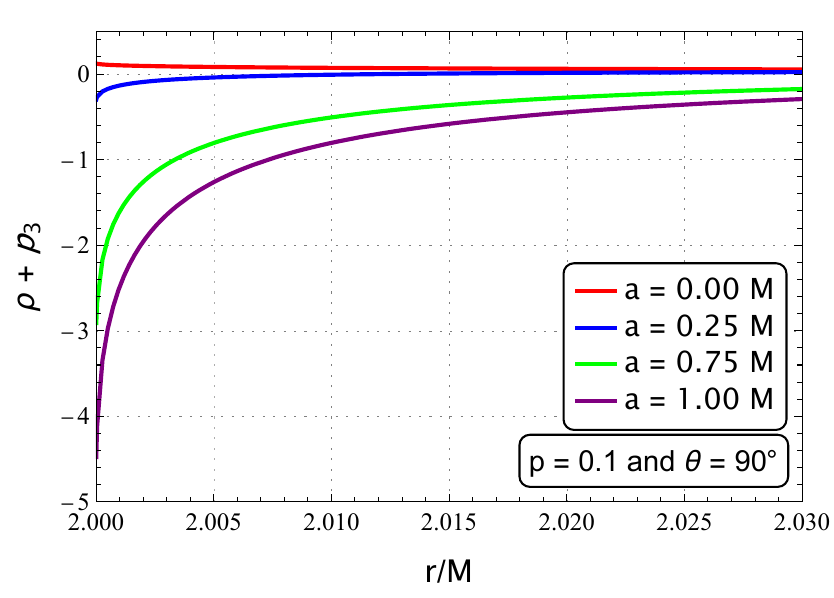}
         \includegraphics[width=0.326\textwidth]{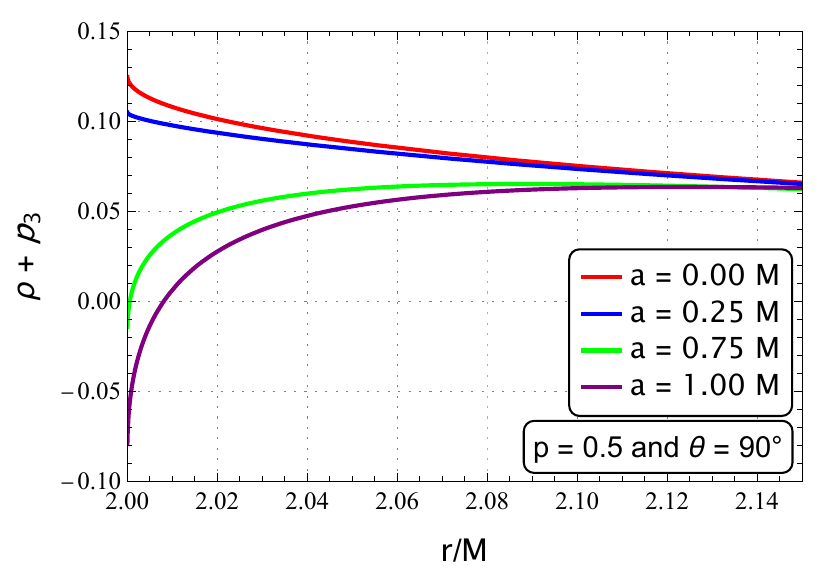}
         \includegraphics[width=0.326\textwidth]{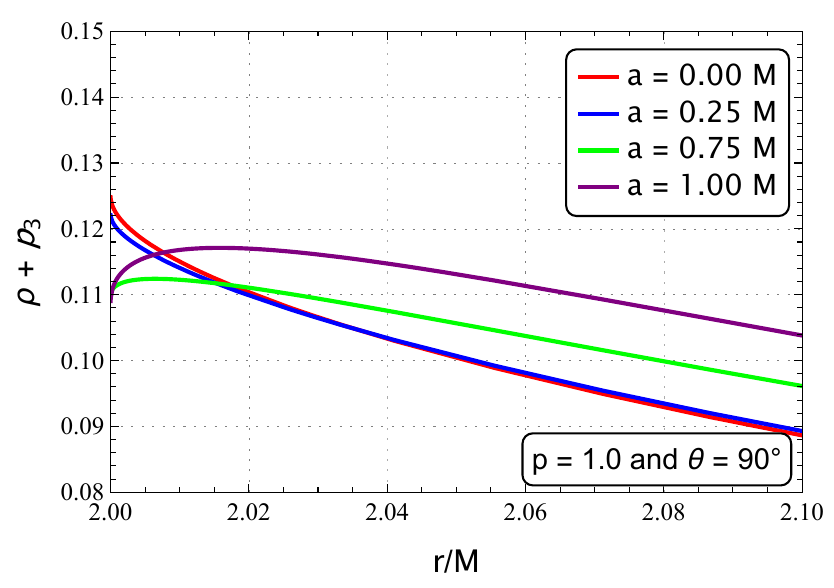}
        \caption{The variation of different null energy conditions are shown w.r.t. $r$ for different values of $a$ and $p$ at $\theta=\pi/2$.}
        \label{fig:NEC}
\end{figure}
\noindent The top row of Figure \ref{fig:NEC} represents the variation of $(\rho+p_{1})$, which is always negative for different values of $a$ and $p$. However, in the plots shown in the last two rows (middle and lowermost) of Figure \ref{fig:NEC}, energy conditions do appear to hold for some values of $a$ and $p$ over a 
finite domain of $r$  (not  for all $r$). Therefore, expectedly, 
since our geometry is a wormhole, the Null Energy Condition
is indeed violated 
for all values of $a$ and $p$. Our result 
is consistent with the previously 
obtained violation of the energy conditions
(in the framework of GR) in the static metric ($a=0$). The violation of the null energy condition further confirms that the other energy conditions will also be violated \cite{Visser}. Thus the required matter to sustain our rotating wormhole geometry violates all energy conditions, as long as we view the geometry as a solution in GR. 
\begin{figure}[h]
         \includegraphics[width=0.326\textwidth]{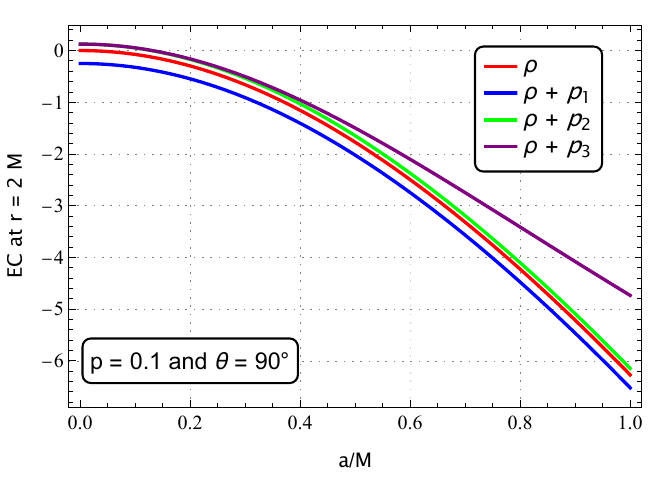}
         \includegraphics[width=0.326\textwidth]{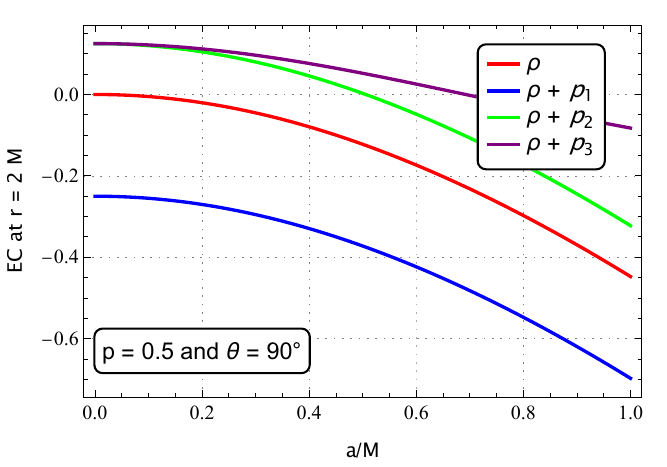}
         \includegraphics[width=0.326\textwidth]{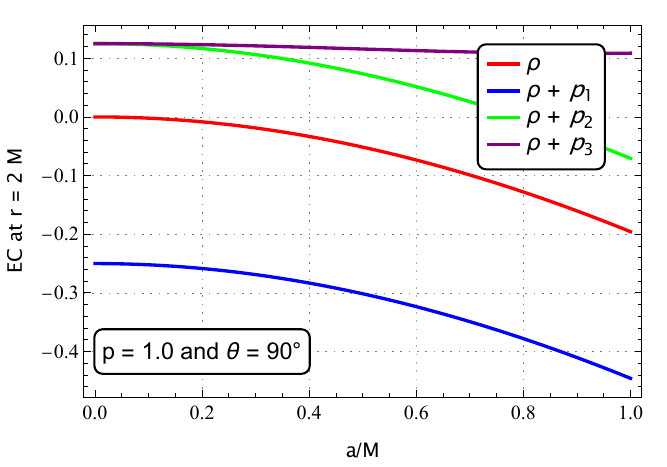}
         \includegraphics[width=0.326\textwidth]{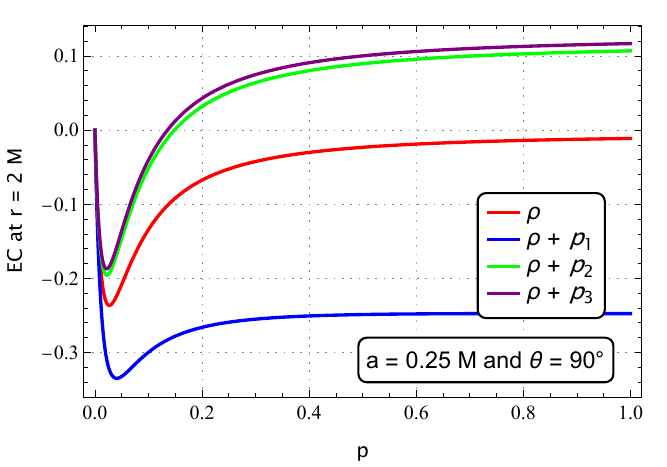}
         \includegraphics[width=0.326\textwidth]{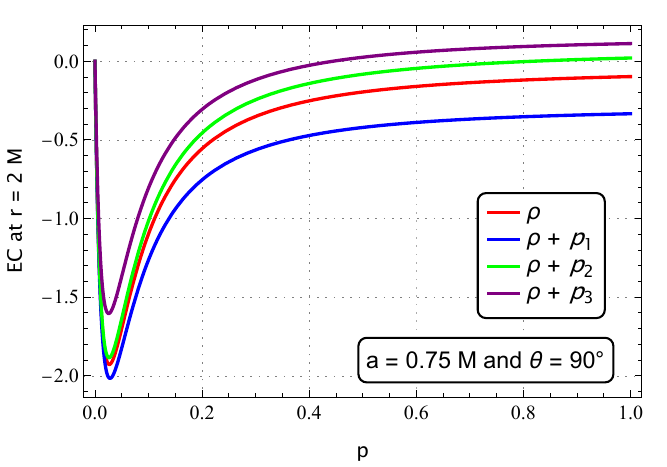}
         \includegraphics[width=0.326\textwidth]{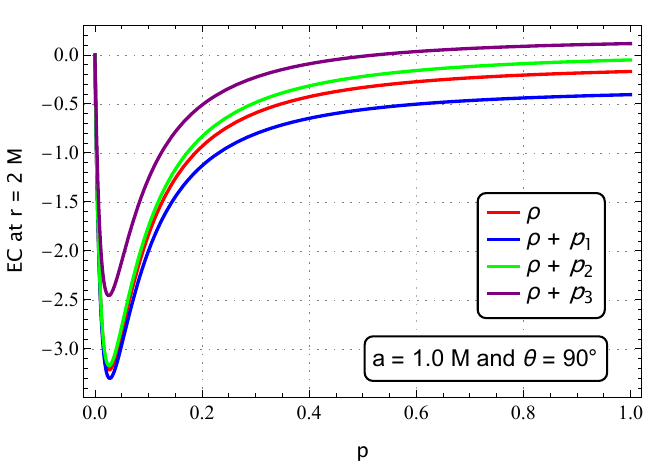}
        \caption{The variation of Null and Weak energy conditions (EC) at the throat is shown w.r.t. metric parameters at $\theta=\pi/2$. (Top panel) for a fixed value of $p$, Null and Weak EC have more negative value with increasing $a$. (Bottom panel) for a fixed value of $a$, Null and Weak EC improves as $p$ increases. Note that $p=0$ represents the vacuum solution.}
        \label{fig:ecatthroat}
\end{figure}

\noindent Next, we have analysed the effects on the energy conditions due to the inclusion of parameters $p$ and $a$ in our metric. We chose the throat radius $(r=2M)$ as our point of analysis since the negativity is maximum there. The variation of different energy conditions at the throat, as a function of $a$ and $p$ appear in Figure \ref{fig:ecatthroat}. It is evident from the top panel of Figure \ref{fig:ecatthroat} that for a fixed value of $p$, as we increase the rotational parameter $a$, the violation of energy conditions worsens. Similarly, the bottom panel of Figure \ref{fig:ecatthroat} shows that the violation of the energy conditions at $r=2M$ is lesser (approaches positive values) as we increase $p$ for a fixed $a$. It is important to note that $p=0$ represents the vacuum Kerr black hole solution where all energy conditions hold trivially. On the contrary, for an arbitrary value of $p$ greater than zero (however small) we have a wormhole where energy conditions do not hold at the throat. Thus, a discontinuity in the energy conditions is expected as $p\to 0$, which is reflected in Figure \ref{fig:ecatthroat} (bottom panel). In summary, a geometry with a relatively larger value of $p$ and a smaller $a$ 
can result in the required matter violating the energy conditions, but to
a comparatively lesser extent. 

\subsection{Curvature invariants and nonsingular behaviour}
\noindent A spacetime solution in any theory of gravity is of special interest when it is non-singular. Smooth, continuous behaviour as well as finite values of all curvature invariants throughout the domains of all coordinates is a necessity for any non-singular geometry. Zakhary and McIntosh (ZM) \cite{Zakhary} (see also
\cite{narlikar} and \cite{carminati} for earlier pioneering work) showed that, in general, seventeen Riemann curvature invariants may be required to characterize a four dimensional spacetime.
Further in \cite{Zakhary}, the authors showed that in addition to the Ricci scalar, a minimally independent set of ZM invariants for a Petrov type D spacetime with Segre type $[(1,1) (1 1)]$ can be formed by the Weyl invariants $I_1$, $I_2$, the Ricci invariant $I_6$, and the mixed invariants $I_9$, $I_{10}$ which are given by
\begin{equation}
\begin{split}
    I_1&= C_{\alpha\beta\mu\nu}C^{\alpha\beta\mu\nu},\quad~ I_2= -C^{\alpha\beta\mu\nu}C^{*}_{\alpha\beta\mu\nu}, \quad~ I_6= R_{\mu\nu}R^{\mu\nu}, \quad~ 
    I_9= C_{\alpha \beta \mu}{}^{\nu} R^{\beta \mu}R_{\nu}{}^{\alpha},\\
    \text{and} & \quad~ I_{10}= -C^{*}{}_{\alpha \beta \mu}{}^{\nu} R^{\beta \mu}R_{\nu}{}^{\alpha},\\
    \end{split}
    \label{eqs:curvature_invariants}
\end{equation}
where $C_{\alpha\beta\mu\nu}$ and $C^{*}_{\alpha\beta\mu\nu}$ are the Weyl tensor and the dual Weyl tensor defined by
\begin{equation}
\begin{split}
    C_{\alpha\beta\mu\nu}&= R_{\alpha\beta\mu\nu} + \frac{R}{6} \left(g_{\alpha\mu}g_{\beta\nu}-  g_{\alpha\nu}g_{\beta\mu}\right) - \frac{1}{2} \left( g_{\alpha\mu}R_{\beta\nu}-  g_{\alpha\nu}R_{\beta\mu} - g_{\beta\mu}R_{\alpha\nu}+  g_{\beta\nu}R_{\alpha\mu}\right),\\
     C^{*}_{\alpha\beta\mu\nu}&= \frac{1}{2} \sqrt{-g} \epsilon_{\alpha \beta \rho \sigma} C^{\rho\sigma}{}_{\mu\nu}. 
\end{split}    
\end{equation}
 All other Riemann curvature invariants can be expressed in terms of these six ZM invariants. The Kretschmann scalar ($K$) can be written as
 $K=R_{\alpha\beta\mu\nu}R^{\alpha\beta\mu\nu}= I_1+ 2I_6 - R^2/3$. 
 
\begin{figure}[!htbp]
\centering
\subfigure[\hspace{0.1cm}$a/M = 1.0$ and $p = 0$]{\includegraphics[width=0.326\textwidth]{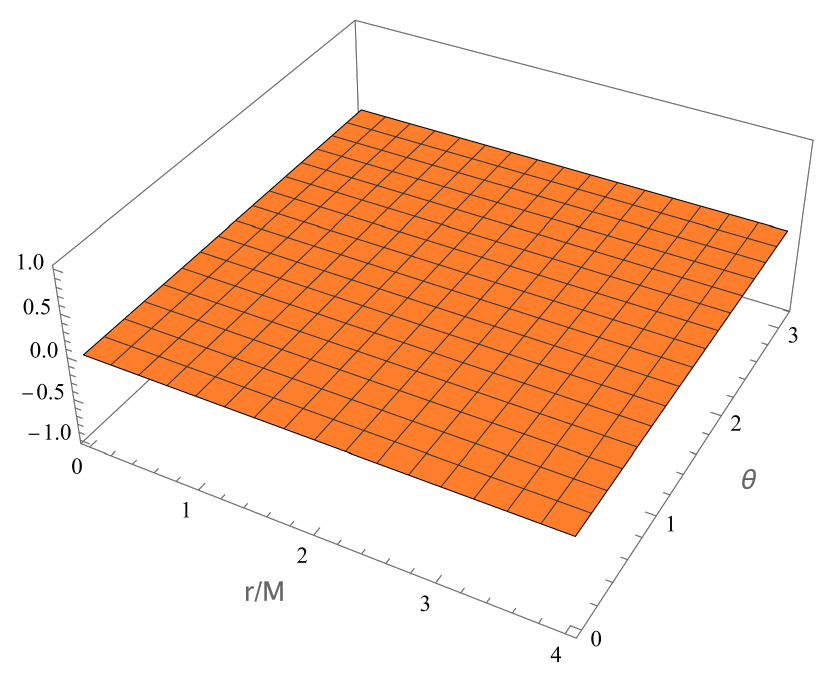}\label{subfig:Ricci1}}
\subfigure[\hspace{0.1cm}$a/M = 0.5$ and $p = 0.5$]{\includegraphics[width=0.326\textwidth]{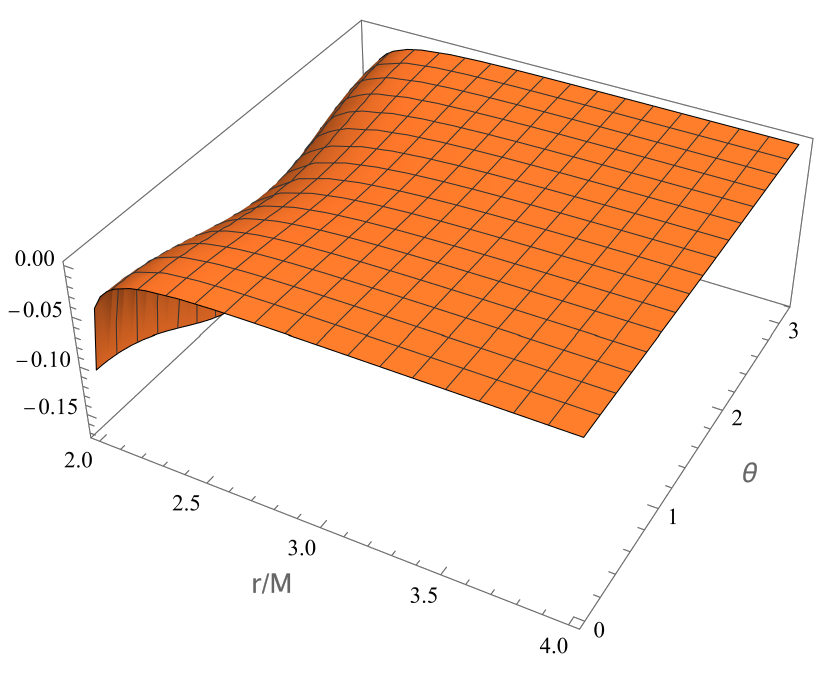}\label{subfig:Ricci2}}
\subfigure[\hspace{0.1cm}$a/M = 0$ and $p = 1.0$]{\includegraphics[width=0.326\textwidth]{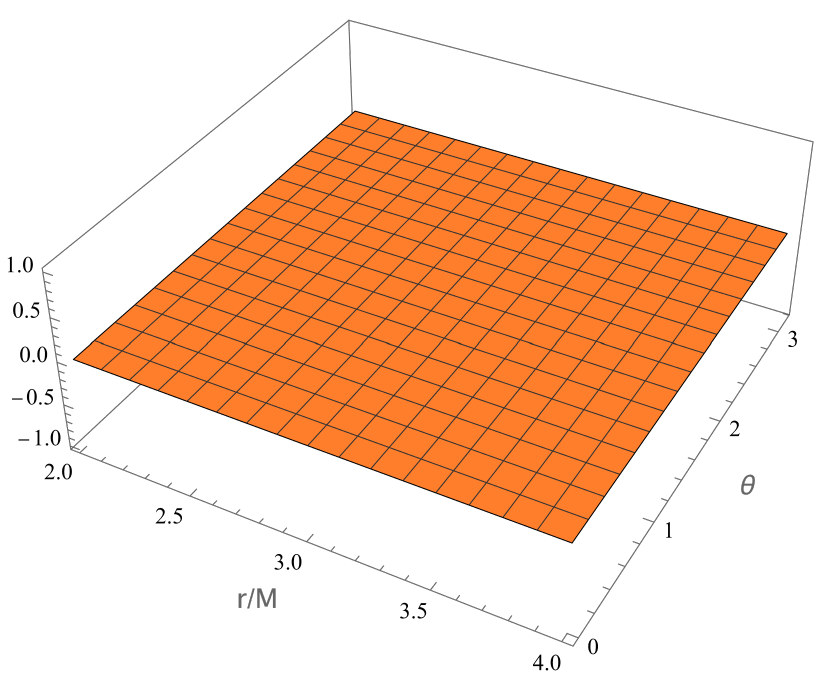}\label{subfig:Ricci3}}
\caption{Variation of Ricci scalar ($g^{\mu\nu}R_{\mu\nu}$) with $r$ and $\theta$ for different sets of $a$ and $p$}
\label{fig:WH_Ricci scalar}
\end{figure}

\begin{figure}[!htbp]
\centering
\subfigure[\hspace{0.1cm}$a/M = 1.0$ and $p = 0$]{\includegraphics[width=0.326\textwidth]{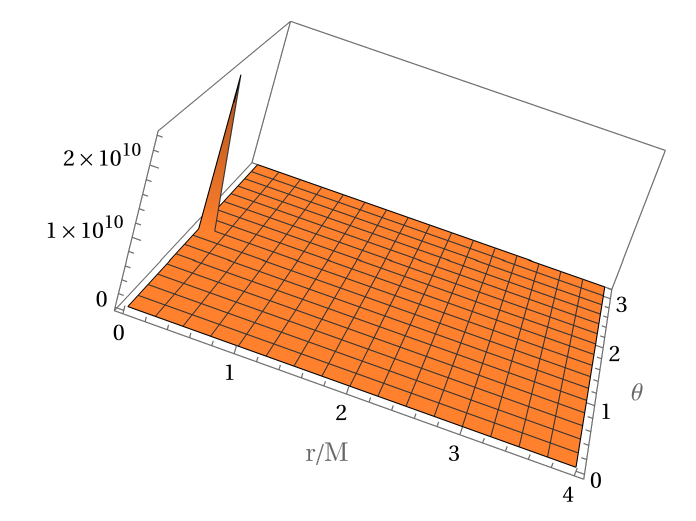}\label{subfig:Riccisq1}}
\subfigure[\hspace{0.1cm}$a/M = 0.5$ and $p = 0.5$]{\includegraphics[width=0.326\textwidth]{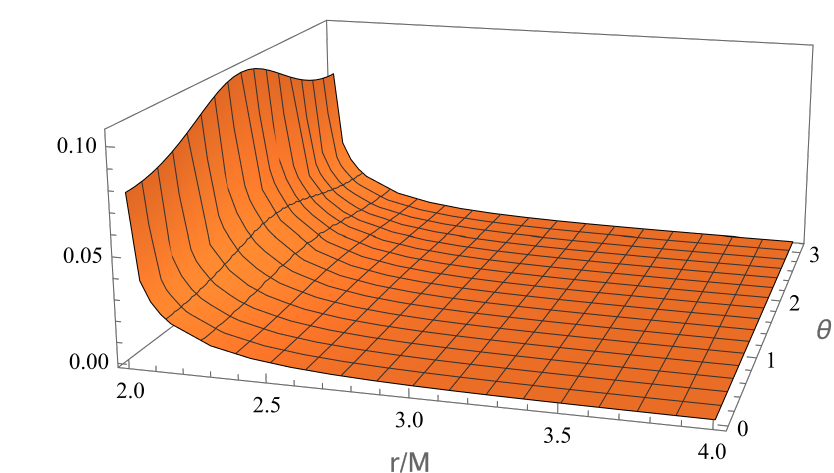}\label{subfig:Riccisq2}}
\subfigure[\hspace{0.1cm}$a/M = 0$ and $p = 1.0$]{\includegraphics[width=0.326\textwidth]{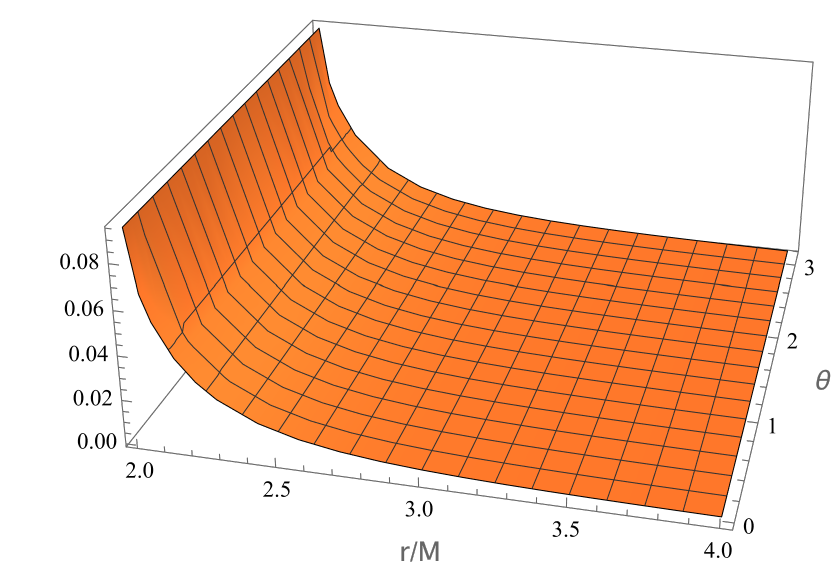}\label{subfig:Riccisq3}}
\caption{Variation of  $R^{\mu\nu}R_{\mu\nu}$ scalar with $r$ and $\theta$ for different sets of $a$ and $p$}
\label{fig:Ricci square scalar}
\end{figure}

\begin{figure}[!htbp]
\centering
\subfigure[\hspace{0.1cm}$a/M = 1.0$ and $p = 0$]{\includegraphics[width=0.326\textwidth]{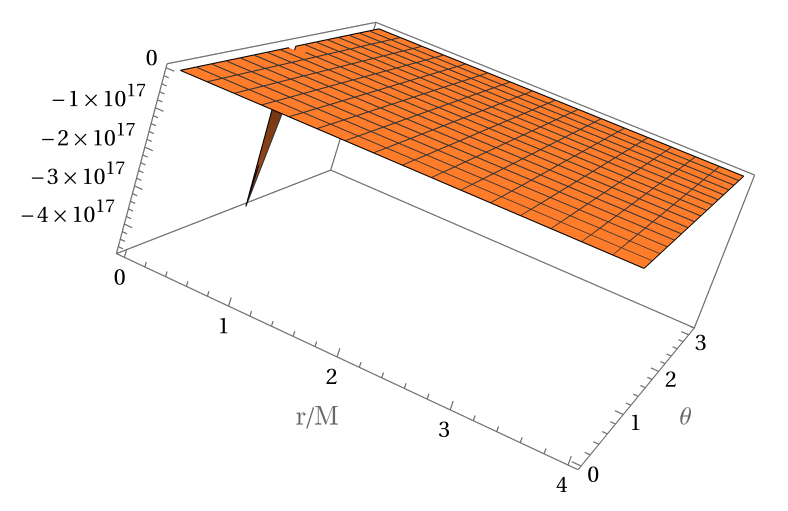}\label{subfig:Weyl1}}
\subfigure[\hspace{0.1cm}$a/M = 0.5$ and $p = 0.5$]{\includegraphics[width=0.326\textwidth]{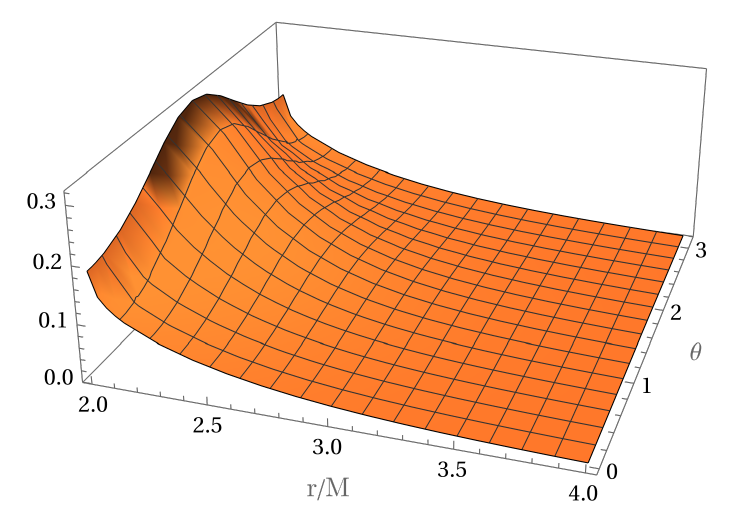}\label{subfig:Weyl2}}
\subfigure[\hspace{0.1cm}$a/M = 0$ and $p = 1.0$]{\includegraphics[width=0.326\textwidth]{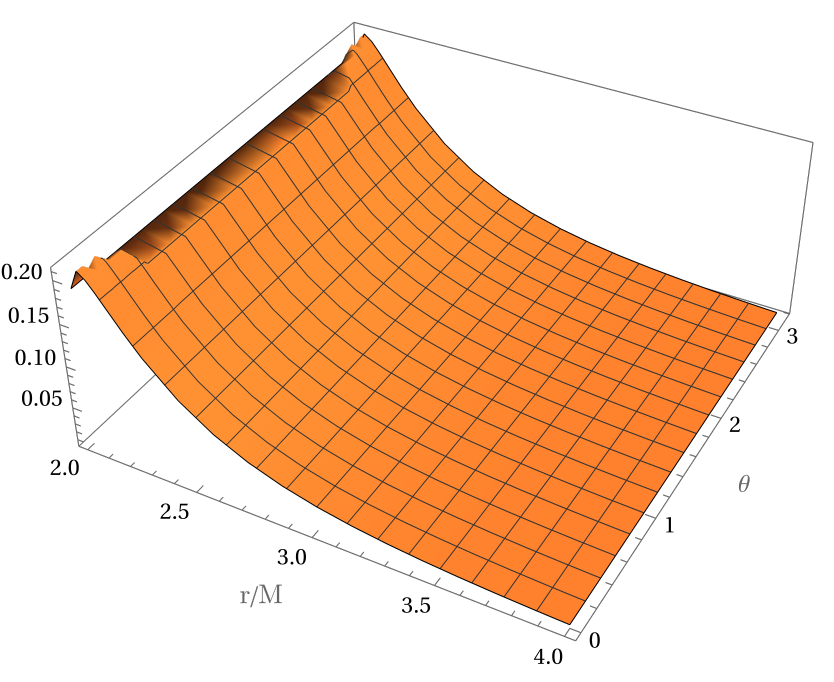}\label{subfig:Weyl3}}
\caption{Variation of  $C^{\mu\nu\lambda\delta}C_{\mu\nu\lambda\delta}$ scalar with $r$ and $\theta$ for different sets of $a$ and $p$}
\label{fig: Weyl square scalar}
\end{figure}

\begin{figure}[!htbp]
\centering
\subfigure[\hspace{0.1cm}$a/M = 1.0$ and $p = 0$]{\includegraphics[width=0.326\textwidth]{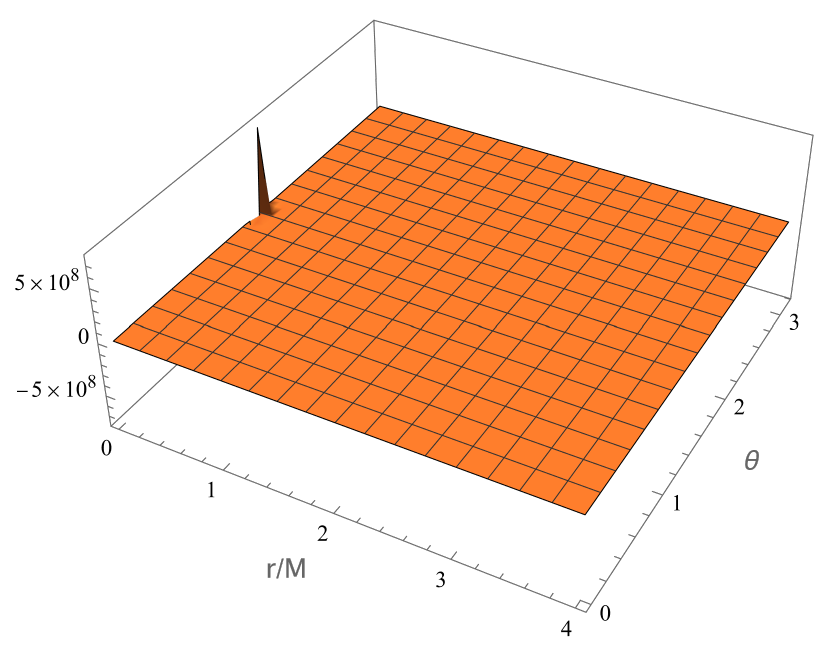}\label{subfig:I2_1}}
\subfigure[\hspace{0.1cm}$a/M = 0.5$ and $p = 0.5$]{\includegraphics[width=0.326\textwidth]{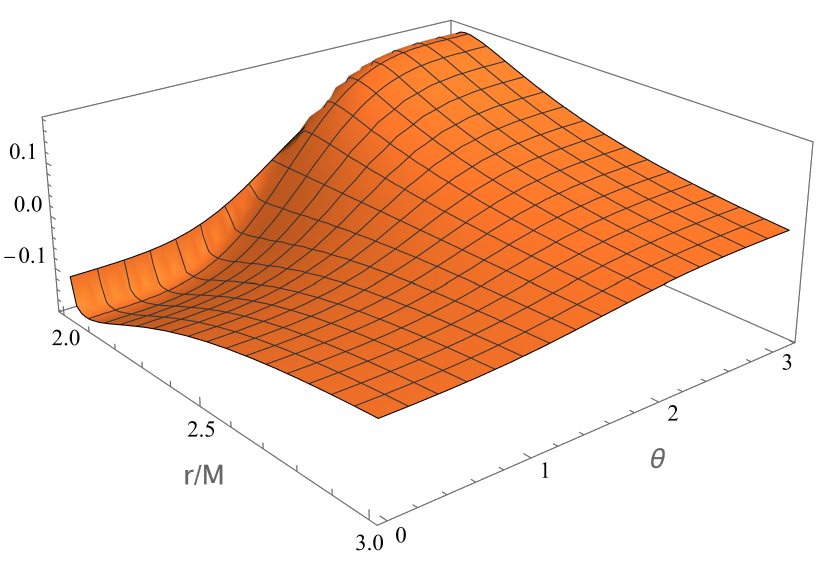}\label{subfig:I2_2}}
\subfigure[\hspace{0.1cm}$a/M = 0$ and $p = 1.0$]{\includegraphics[width=0.326\textwidth]{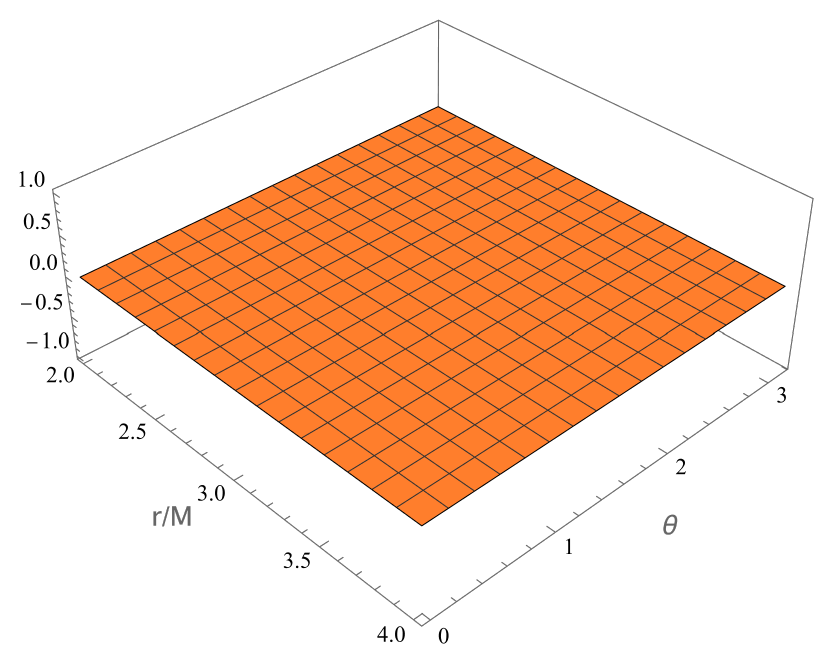}\label{subfig:I2_3}}
\caption{Variation of the invariant $I_2$ with $r$ and $\theta$ for different sets of $a$ and $p$}
\label{fig: I2 scalar}
\end{figure}

\begin{figure}[!htbp]
\centering
\subfigure[\hspace{0.1cm}$a/M = 1.0$ and $p = 0$]{\includegraphics[width=0.326\textwidth]{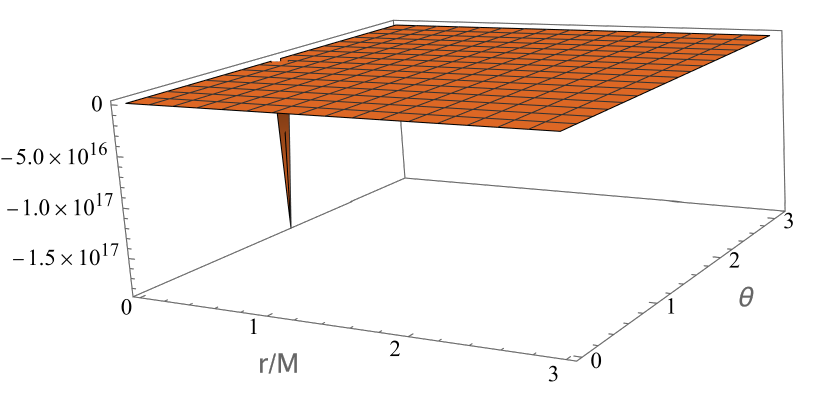}\label{subfig:I9_1}}
\subfigure[\hspace{0.1cm}$a/M = 0.5$ and $p = 0.5$]{\includegraphics[width=0.326\textwidth]{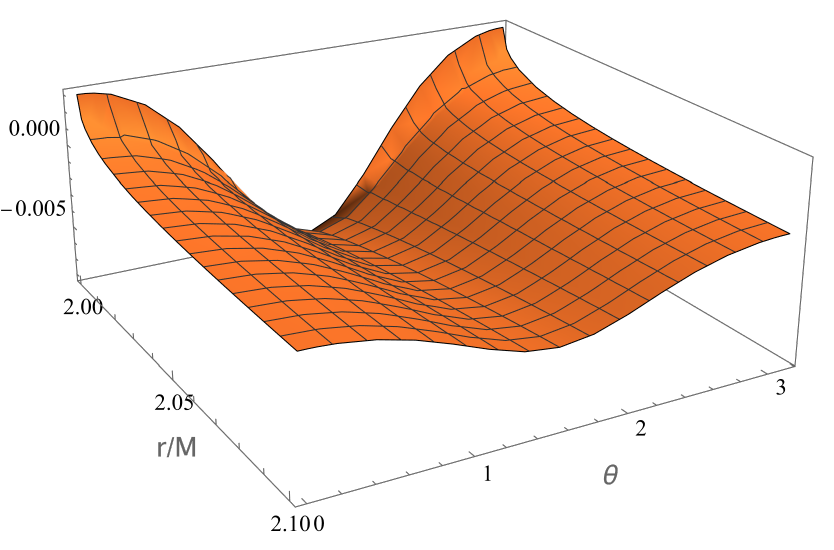}\label{subfig:I9_2}}
\subfigure[\hspace{0.1cm}$a/M = 0$ and $p = 1.0$]{\includegraphics[width=0.326\textwidth]{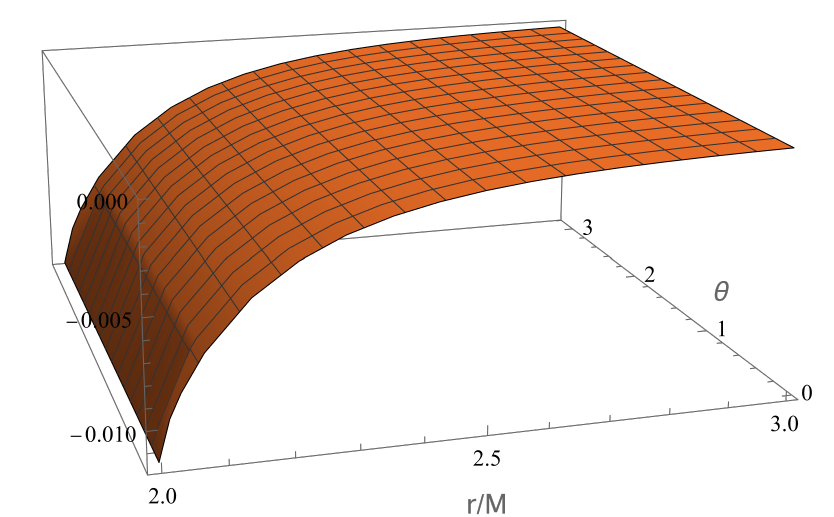}\label{subfig:I9_3}}
\caption{Variation of the invariant $I_9$ with $r$ and $\theta$ for different sets of $a$ and $p$}
\label{fig: I9 scalar}
\end{figure}

\begin{figure}[!htbp]
\centering
\subfigure[\hspace{0.1cm}$a/M = 1.0$ and $p = 0$]{\includegraphics[width=0.326\textwidth]{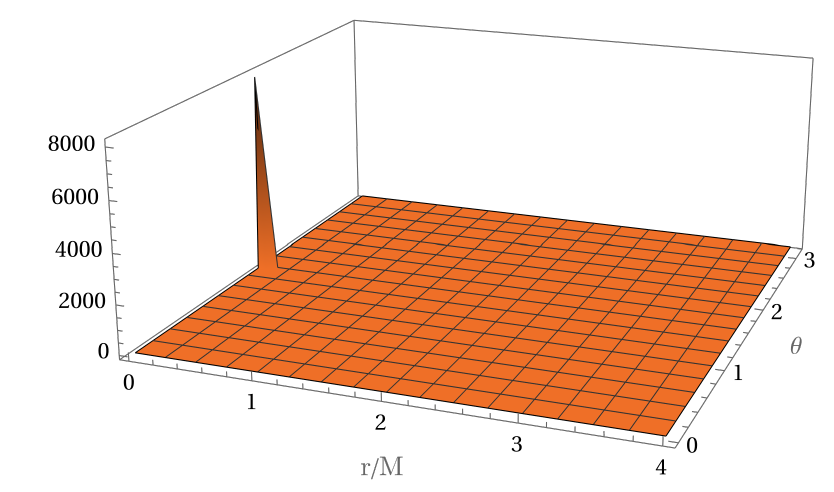}\label{subfig:I10_1}}
\subfigure[\hspace{0.1cm}$a/M = 0.5$ and $p = 0.5$]{\includegraphics[width=0.326\textwidth]{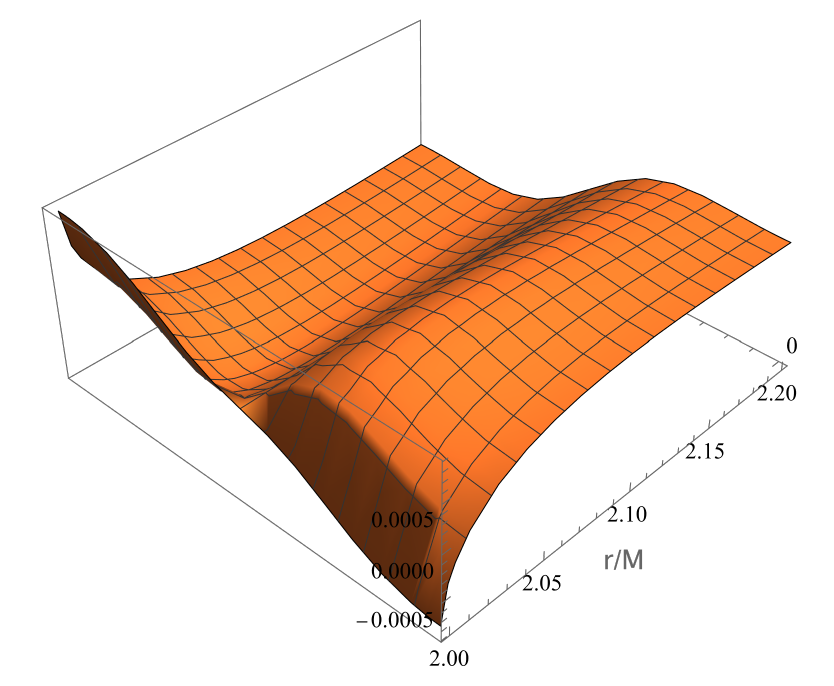}\label{subfig:I10_2}}
\subfigure[\hspace{0.1cm}$a/M = 0$ and $p = 1.0$]{\includegraphics[width=0.326\textwidth]{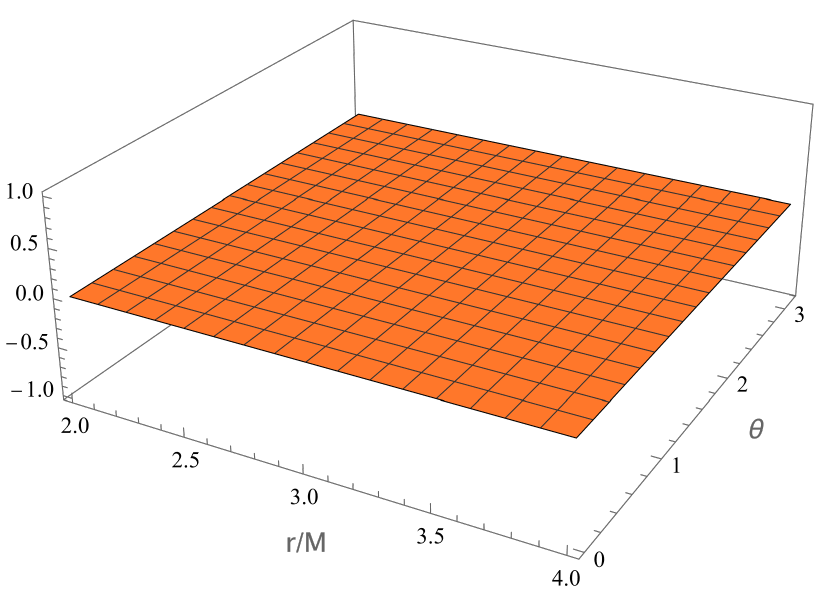}\label{subfig:I10_3}}
\caption{Variation of the invariant $I_{10}$ with $r$ and $\theta$ for different sets of $a$ and $p$}
\label{fig: I10 scalar}
\end{figure}

\noindent As proved in \cite{Torres}, the particular metric structure of our rotating wormhole in Eq.(\ref{eq:rotating_wormhole_new}) enables it to be a Petrov type D spacetime and Segre type $[(1,1) (1 1)]$. Therefore, to demonstrate the non-singular character of our rotating wormhole, we present the behaviour of six curvature invariants (including the Ricci scalar) in the coordinate space $(r/M,\theta)$ for different parameter values, in Figures \ref{fig:WH_Ricci scalar}, \ref{fig:Ricci square scalar}, \ref{fig: Weyl square scalar} ,\ref{fig: I2 scalar}, \ref{fig: I9 scalar}, and \ref{fig: I10 scalar}.
It is important to note that $r\in [2M,\infty)$ and $\theta\in[0,\pi]$, except for the Kerr case ($p=0$) where $r\in[0,\infty)$.

\noindent Fig. \ref{subfig:Ricci1} demonstrates the zero value of the Ricci scalar for all values of $r$ and $\theta$ where $a=M$ and $p=0$, i.e. Kerr metric, which is expected. Similarly, $a=0$ represents the static $R=0$ wormhole geometry, which is represented in Fig \ref{subfig:Ricci3}. However, a finite value of $a$ and $p$ leads to a non-zero value of the Ricci scalar (Fig. \ref{subfig:Ricci2}). Thus, in general, the rotating version of the $R=0$ geometry constructed by us has a finite, non-zero  Ricci scalar.

\noindent In the other figures (Figs. 6-10), we have shown the smooth and finite behaviour of the other independent curvature invariants as defined in Eqs.~\ref{eqs:curvature_invariants} for the entire domain of the radial $(r)$ and angular $(\theta)$ coordinates. One may note that when $a\neq 0$ and $p=0$, i.e. the Kerr metric, all the curvature scalar invariants diverge at $r\to 0,\theta=\pi/2$. Therefore, the above-discussed rotating wormhole solution is non-singular in nature $(p\neq 0)$.

\section{The shadow of the rotating wormhole}
\noindent Let us now address the issue of relevance of the above-discussed wormhole in the context of observational astrophysics. To detect the existence of any wormhole, one needs to propose signatures 
related to different phenomena which can be tested with available
observations. Such phenomena include light deflection, orbits, time delay, etc. In this article, we choose to analyze the shadow of the above-discussed rotating wormhole and try to restrict the metric parameters from available observational data.

\noindent To begin our analysis, we consider the usual assumptions made 
while calculating the wormhole shadow as previously mentioned by different authors \cite{Shaikh, Nedkova, Rahaman}. In a wormhole geometry, which connects two different regions of spacetime, all the light sources are in one region, and the other is free from light near the throat. Therefore, in the illuminated region, light follows two possible trajectories around the wormhole throat. Either photons can enter  the wormhole and pass through the throat or they may be scattered by the wormhole and reach infinity. An asymptotic observer in the illuminated region can receive the scattered photon and consider it as a bright point in his/her sky. Photons that fall into the wormhole will appear as a dark spot in the observer's sky. This dark region on the bright background is known as the wormhole shadow. To determine the boundary of the dark region, we have to find out the critical photon orbits that differentiate the infalling and scattered photons by analysing the null geodesics of the wormhole geometry. The trajectories of photons (null geodesics) can be determined by analysing the Hamilton-Jacobi (HJ) equation \cite{Konoplya}
\begin{equation}\label{5.1}
    H(x^{\mu},p_{\mu})+\frac{\partial S}{\partial \lambda}=0
\end{equation}
where $\lambda$ is an affine parameter, $S$ is the Jacobi action and $H$ is the Hamiltonian corresponding to the null geodesic. As our rotating wormhole geometry is a stationary axisymmetric spacetime, there are two constants of motion -- the energy of the particle $E$ and the angular momentum $L$ about the axis of symmetry (metric is independent of $t$ and $\phi$).  An additional symmetry leads to another conserved quantity (Carter constant \cite{Carter}), which makes the solution of the HJ equation separable. The separable solution can be written in the following form,
\begin{equation}\label{5.2}
    S=-Et+L\phi+S_{r}(r)+S_{\theta}(\theta)
\end{equation}
where $E=-\frac{\partial S}{\partial t}$, $L=\frac{\partial S}{\partial \phi}$, and $S_{r}(r)$ and $S_{\theta}(\theta)$ are functions of respective coordinates. As the expression of $S$ (Eqn. \ref{5.2}) is independent of the affine parameter $\lambda$, the HJ equation for null geodesics becomes,
\begin{equation}
    H=\frac{1}{2}g^{\mu\nu}p_{\mu}p_{\nu}=0
\end{equation}
where $p_{\mu}=\frac{\partial S}{\partial x^{\mu}}$ represents the conjugate momentum. We simplify the above equation and obtain the separated angular and radial equations for the functions $S_{r}(r)$ and $S_{\theta}(\theta)$ \cite{Konoplya1}. The angular equation of motion is
\begin{equation}\label{5.4}
    \mathit{C}=\left(\frac{dS_{\theta}}{d\theta}\right)^{2}+\cot^{2}\theta L^{2}-\cos^{2}\theta E^{2}
\end{equation}
where $\mathit{C}$ is the Carter constant. The radial equation is
\begin{equation}\label{5.5}
    r^{2}\Delta(r)\left(1-\frac{2m(r)}{r}+\frac{a^{2}}{r^{2}}\right)\left(\frac{dS_{r}}{dr}\right)^{2}=V(r)
\end{equation}
Here, the effective potential $V(r)$ has the following form
\begin{equation}\label{5.6}
    V(r)=\left(r^{2}E+a^{2}E-aL\right)^{2}-r^{2}\mathit{\kappa}\left(1-\frac{2m(r)}{r}+\frac{a^{2}}{r^{2}}\right)
\end{equation}
where $\mathit{\kappa}\equiv\mathit{C}+\left(L-aE\right)^{2} > 0$.

\noindent The $r\theta$ separability of the HJ
equation implies the existence of a Killing tensor $K^{ij}$ which generates the Carter constant \cite{killing1,killing2,killing3} as a constant of
motion. The essential result follows from the identification of the
inverse metric in the form:
\begin{eqnarray}
g^{ij} = \frac{1}{f_\xi (\xi) - f_\eta(\eta)}\left ( {X^{ij} (\xi)+ Y^{ij} (\eta)} \right )
\end{eqnarray}
The Killing tensor can therefore be shown to be given as:
\begin{eqnarray}
    K^{ij} = -\frac{1}{f_\xi (\xi) - f_\eta(\eta)} \left ({f_\xi \, Y^{ij} (\eta)+ f_\eta \, X^{ij} (\xi)} \right )
\end{eqnarray}
where $X^{\eta j} = Y^{\xi j}=0$ and $\xi$, $\eta$ are identified with $r$, $\theta$
for Kerr or our metric. One needs to read off $X^{ij}$ and $Y^{ij}$, $f_\xi$ and
$f_\eta$ from the metric functions directly. As shown in 
\cite{killing1,killing2} and stated in \cite{killing3} 
the above result follows from the separability of the HJ equation and it is simple to apply it to our spacetime. It may further be noted, that the
above result is applicable to any axially symmetric rotating metric
where a $\Delta(r)$ and a
$m(r)$ are specified.  This includes the metrics in \cite{killing3} which
obey certain constraints on the $X^{ij}$, $Y^{ij}$ as well as the metric discussed in this article which
belongs to the same general class mentioned in \cite{killing3}. We have not 
explored Klein-Gordon separability for our metric yet, though we hope to discuss it in future work.

\subsection {Shadow profile}
\begin{figure}[h]
         \includegraphics[width=0.326\textwidth]{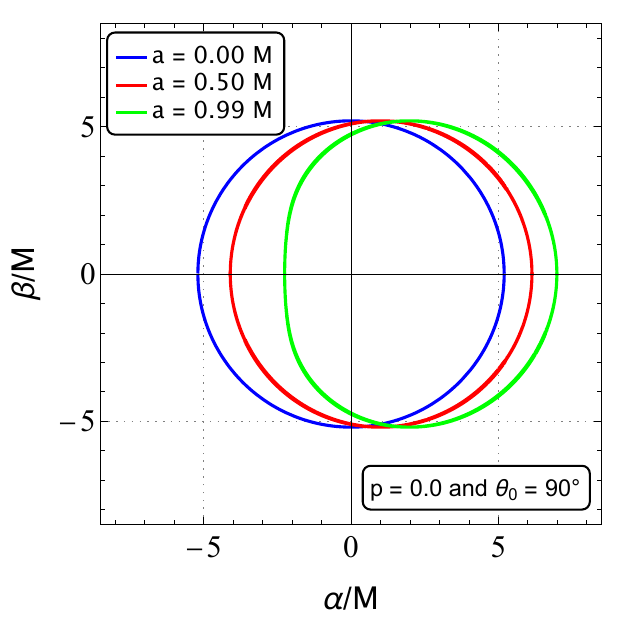}
         \includegraphics[width=0.326\textwidth]{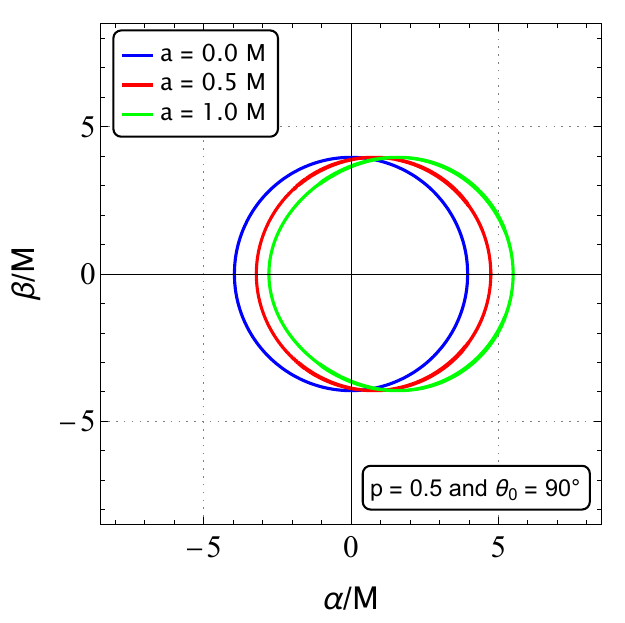}
         \includegraphics[width=0.326\textwidth]{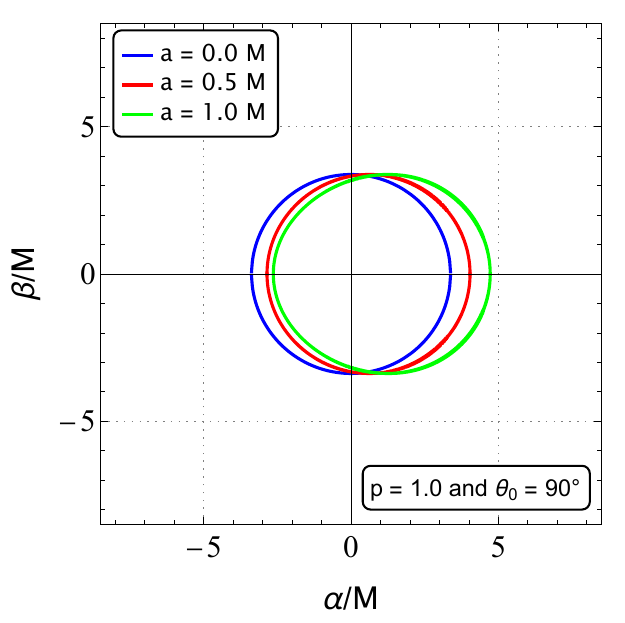}
         \includegraphics[width=0.326\textwidth]{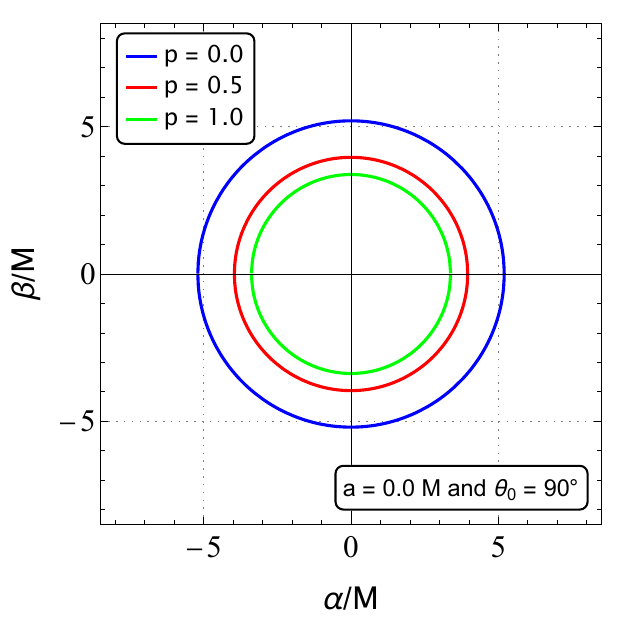}
         \includegraphics[width=0.326\textwidth]{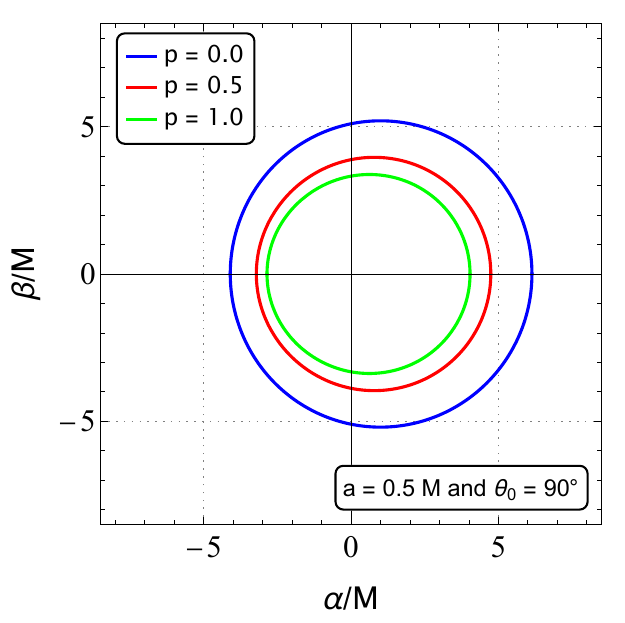}
         \includegraphics[width=0.326\textwidth]{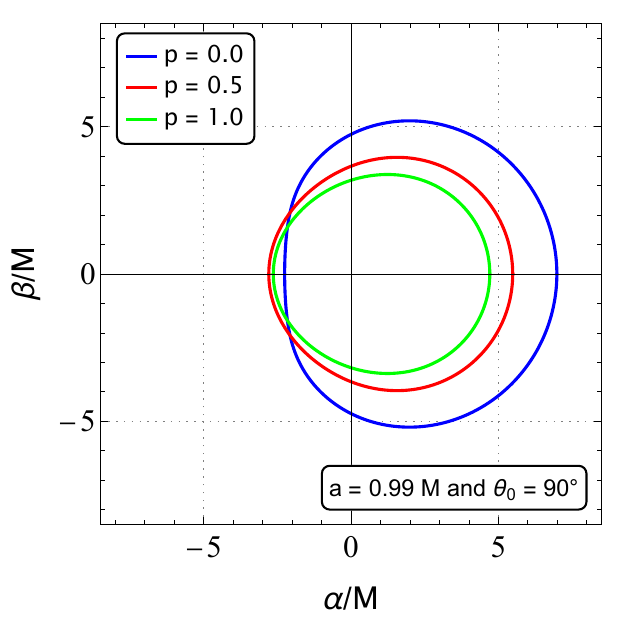}
        \caption{Shadow profiles of the rotating wormhole for different values of $a$ and $p$ at the inclination angle $\theta_{0}=90\degree$. Note that at $p=0.0$, the shadow is the same as Kerr metric}
        \label{fig:Whsh}
\end{figure}

\begin{figure}[h]
         \includegraphics[width=0.326\textwidth]{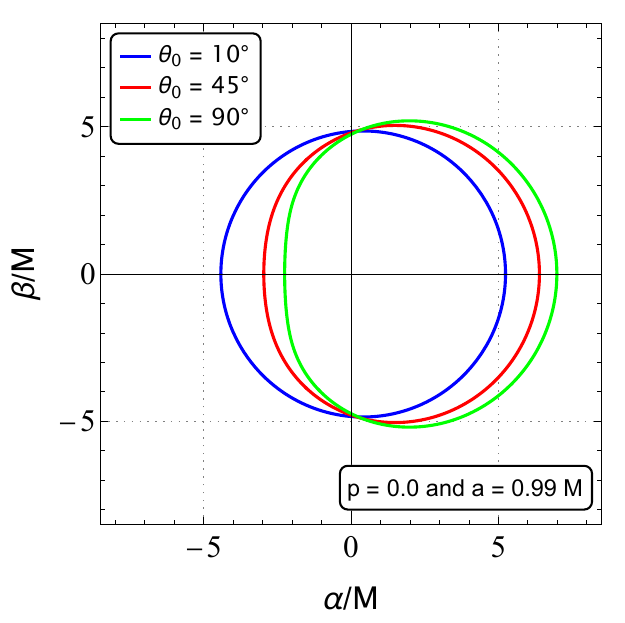}
         \includegraphics[width=0.326\textwidth]{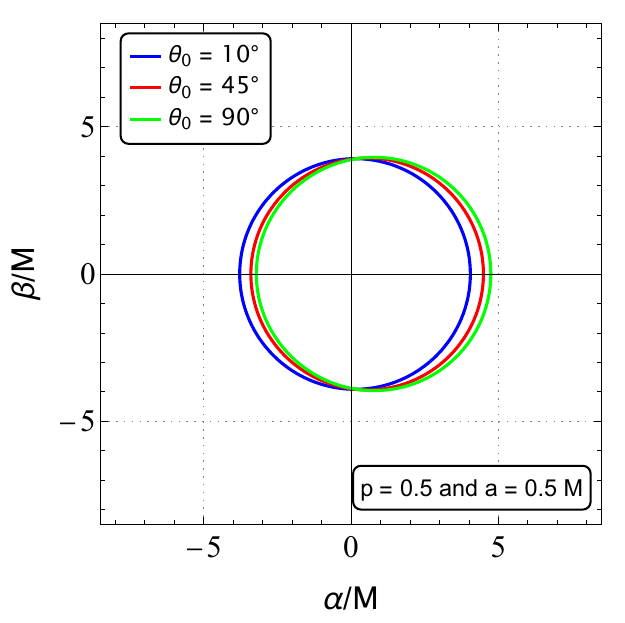}
         \includegraphics[width=0.326\textwidth]{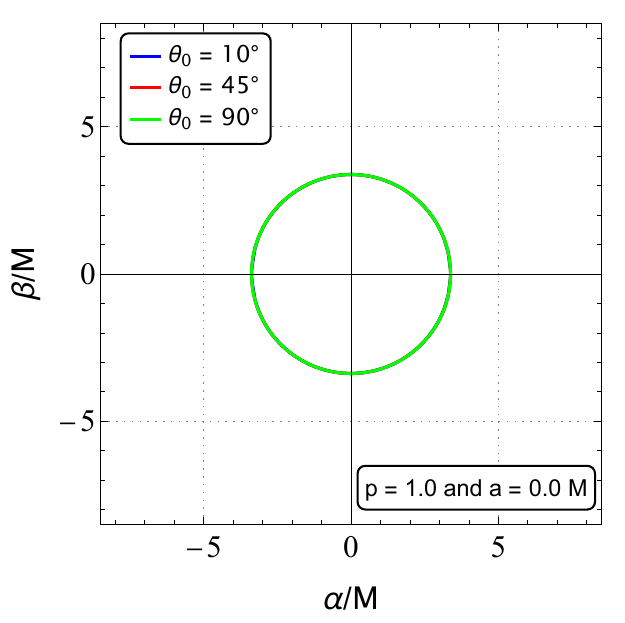}
        \caption{Shadow profiles for different inclination angle $\theta_{0}$ at different $a$, $p$}
        \label{fig:Whsh1}
\end{figure}

\noindent As mentioned previously, to determine the boundary of the shadow, we have to figure out the critical photon orbits that divide the null geodesics into the prescribed two categories (either falling into the wormhole or being scattered to infinity). As the null geodesics are governed by the effective potential described in Eq. (\ref{5.6}), the boundary of the two classes of photon orbits can be obtained as the unstable spherical orbits $(r_{c})$ satisfying the following conditions,
\begin{equation}
    V(r_c)=0, \hspace{1cm} V^{\prime}(r_{c})=0 \hspace{1cm} V^{\prime\prime}(r_c)>0
\end{equation}
where $\{\prime\}$ denotes the radial derivative. Note that an outward perturbation on the photons in the unstable orbits may lead them to reach the asymptotic observer. By solving the above equations, we have obtained the following critical parameters,
\begin{equation}
         \frac{L}{E}=\frac{1}{a}\left(r_{c}^{2}+a^{2}-\frac{4r_{c}^{2}R(r_c)}{2R(r_c)+r_cR^{\prime}(r_c)}\right); \hspace{0.6cm}
    \frac{\mathit{\kappa}}{E^2}=\frac{16r_{c}^{2}R(r_c)}{\left(2R(r_c)+r_cR^{\prime}(r_c)\right)^2}
\end{equation}
where we define $R\equiv \left(1-\frac{2m(r)}{r}+\frac{a^{2}}{r^{2}}\right)$. Although these critical parameters define the boundary of the wormhole shadow, the asymptotic observer will see the projection of it in his/her sky, i.e., the plane passing through the wormhole and normal to the line connecting it with the observer. The coordinates in this plane are known as celestial coordinates denoted as $\alpha$ and $\beta$, constructed by Bardeen \cite{bardeenshadow}.
\begin{equation}\label{5.9}
    \alpha=-r_{0}^{2}\sin{\theta_{0}}\frac{d\phi}{dr}; \hspace{1cm} \beta=r_{0}^{2}\frac{d\theta}{dr}
\end{equation}
where $r_{0}$ is the distance between the wormhole and the observer, and $\theta_{0}$ is the inclination angle between the wormhole rotation axis and the line of sight between the source and observer. Substituting (\ref{5.4}) and (\ref{5.5}) in Eq. (\ref{5.9}) and by taking the limit $r\to \infty$, we have 
\begin{equation}
    \alpha(r_c)=-\frac{1}{\sin\theta_0}\frac{L}{E}; \hspace{1cm} \beta^2(r_c)=\frac{\mathit{C}-\cot^2\theta_0 L^2+\cos^{2}\theta_0 E^2}{E^2}
\end{equation}
The parametric plot of $\alpha(r_c)$ and $\beta(r_c)$ will give us the shadow profile in the observer's sky, which depends on $\theta_0$ as well as the specific values of metric parameters $M$, $a$ and $p$. In Figure \ref{fig:Whsh}, we present the shadow profiles for different values of $a$ and $p$ at the inclination angle $\theta_{0}=90\degree$.
\noindent It is prominent from the upper panel of Figure \ref{fig:Whsh}, that for a fixed value of $p$, an increase in the rotation parameter $a$, leads to a shift of the shadow profile towards the right, in the observers sky. The lower panel of Figure \ref{fig:Whsh} shows how the shape of the shadow changes under the variation of $p$ for a fixed value of $a$. Note that we have the standard results of the Kerr metric when $p=0$. Figure \ref{fig:Whsh1} represents the variation of the shadow profiles under the change in the asymptotic observer's position $(\theta_0)$ for a fixed value of the metric parameters.
It shows that a higher inclination angle leads to a shift in the shadow profile towards the right in the observer's sky.

\subsection{Matching with observations}
\begin{figure}[h]
         \includegraphics[width=0.492\textwidth]{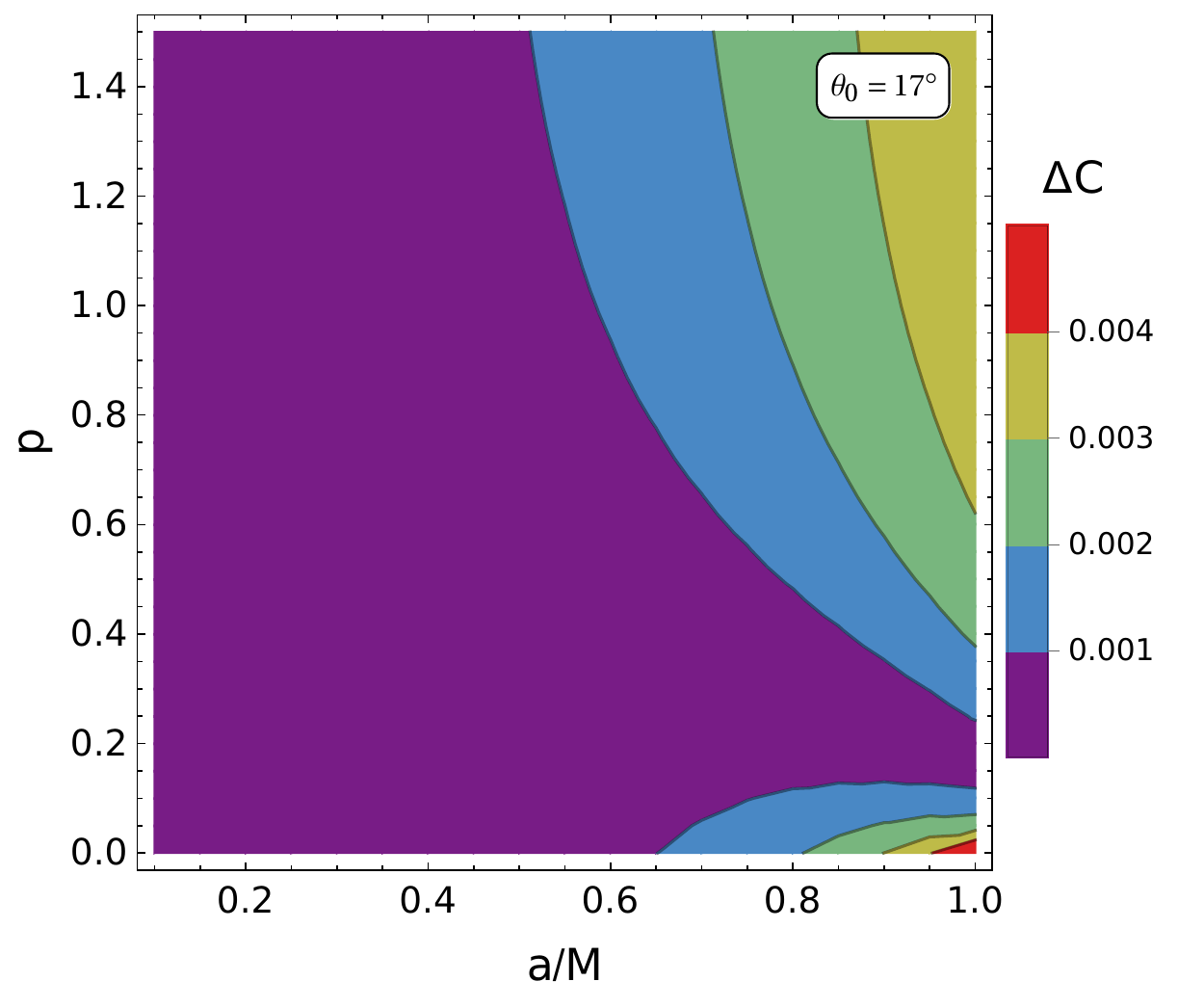}
         \includegraphics[width=0.492\textwidth]{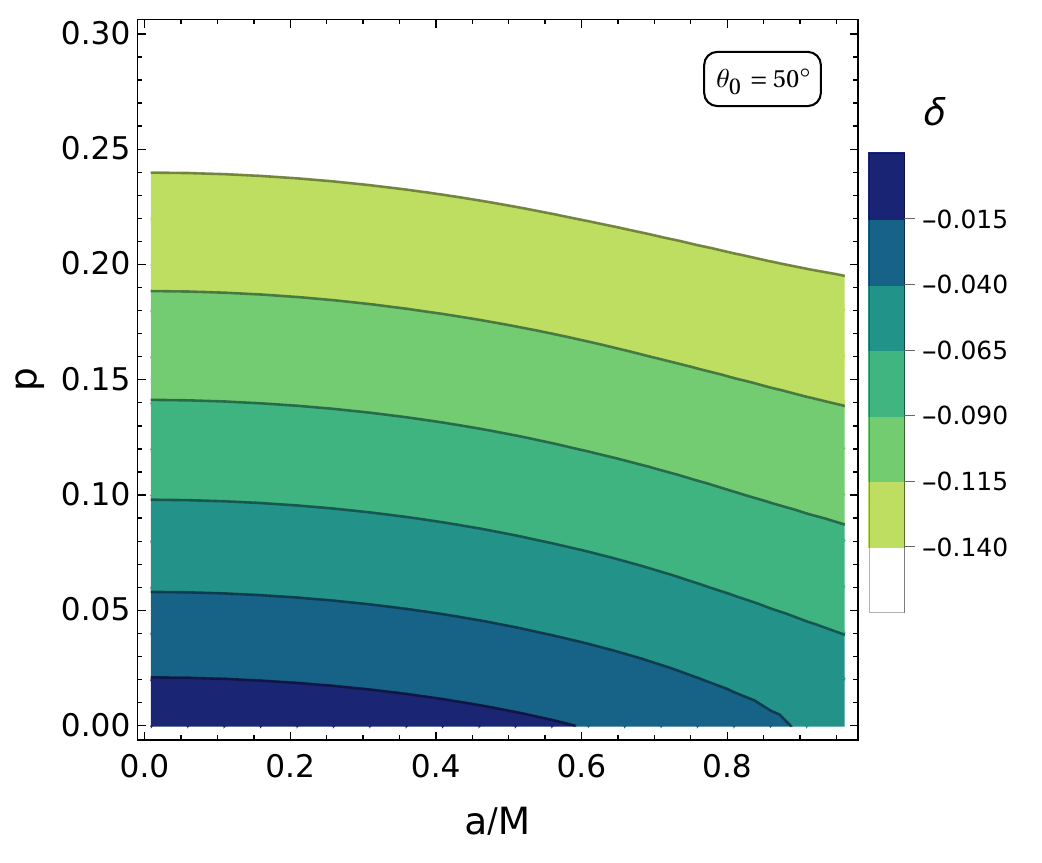}
        \caption{The contour plot for different values of $\Delta C$ at $\theta_0=17\degree$ (left) and $\delta$ at $\theta_0=50\degree$ (right) are shown over the parameter space $a/M$ vs $p$.}
        \label{fig:Constraint}
\end{figure}

\noindent  We now impose restrictions on our wormhole metric parameters $a$ and $p$ using the EHT observations of M$87^*$ and SgrA$^*$ \cite{Akiyama1, Akiyama5, Akiyama6, Akiyama12, Akiyama17}. For this, we will use the standard variables -- deviation from circularity $(\Delta C)$ \cite{Bambi} and fractional deviation parameter $(\delta)$ \cite{Afrin, Shaikh2}, which have bounds imposed by the EHT collaboration. As the shadow profiles have reflection symmetry around $\alpha$-axis, the geometric center $(\alpha_c,\beta_c)$ can be defined by $\alpha_c = \frac{1}{A}\int \alpha\, d A$ and $\beta_c =0$, where $d A = 2\beta d\alpha$ is the area element on the shadow. We also have to consider the angle $\phi$ between the $\alpha$-axis and the vector connecting $(\alpha_c,\beta_c)$ to any arbitrary point on the shadow boundary. Thus, the average shadow radius can be defined as  
\begin{equation}
    R_{avg}^2= \frac{1}{2\pi} \int^{2\pi}_0 d \phi \, \,\ell^2(\phi).
\end{equation}
where $ \ell(\phi)= \sqrt{\left(\alpha(\phi)-\alpha_c\right)^2+\beta(\phi)^2}$. Following \cite{Jana}, the deviation from circularity $(\Delta C)$ can be obtained as 
\begin{equation}
    \Delta C = \frac{1}{R_{avg}}\sqrt{\frac{1}{2\pi}\int^{2\pi}_0 d \phi \left(\ell(\phi)-R_{avg}\right)^2}.
\end{equation}
Thus, $\Delta C$ is the fractional RMS distance from the average shadow radius $(R_{avg})$. Our metric contains three independent parameters -- $M$, $a$ and $p$. Therefore, we constrain the metric parameter space ($a/M$ vs $p$) from the recent EHT observation of M$87^*$ and SgrA$^*$ using $\Delta C$ and $\delta$. From the shadow of M$87^*$, the EHT collaboration concluded that at the inclination angle $(\theta_0)$ $17\degree$, $\Delta C\leq 0.1$ \cite{Akiyama1, Akiyama5, Akiyama6}. However, the restriction on $\Delta C$ from SgrA$^*$ is not available. In Figure \ref{fig:Constraint} (left), $\Delta C$ reaches the maximum value of $0.005$ at an inclination angle $17\degree$ for the parameter values of $p=0$ and $a=0.99$ (Kerr metric). And the full parameter space has $\Delta C < 0.1$. One can extend the parameter space beyond $p>1.5$. However, this will not violate the bound on $\Delta C$. Therefore, we can conclude that all metric parameters are allowed, and we cannot restrict their values from available data on 
the deviation of circularity, i.e. $(\Delta C)$.

\noindent In 2022, the EHT observation on SgrA$^*$ incorporated the fractional deviation parameter $(\delta)$ to compare different spacetime models with Schwarzschild geometry \cite{Akiyama17}. $\delta$ quantifies the deviation of shadow diameter of a geometry from Schwarzschild spacetime and is given by
\begin{equation}
    \delta = \frac{d_{sh}}{d_{sh,Sch}}-1 =\frac{R_{avg}}{3\sqrt{3}M} -1 ,
\end{equation}
where the average diameter of the shadow , $d_{sh}=2R_{avg}$. The EHT collaboration imposed a constraint on $\delta$ by utilising the shadow observations of SgrA$^*$ at an inclination angle $\theta_{0}=50\degree$, together with two distinct sets of previous mass and distance estimates obtained from the VLTI and Keck observations \cite{Akiyama12, Akiyama17,Vagnozzi}. This turns out to be,
\begin{equation}
 \delta= \begin{cases} 
      -0.08^{+0.09}_{-0.09}& \text{(VLTI)} \\
      -0.04^{+0.09}_{-0.10} & \text{(Keck)}  \\
   \end{cases}
\end{equation}
Therefore, at the $1\sigma$ credible level, the fractional deviation parameter varies between $-0.17\leq\delta\leq 0.01$ (VLTI) and $-0.14\leq\delta\leq 0.05$ (Keck). We use the common range of $\delta$, $-0.14<\delta<0.01$ to constrain our rotating wormhole parameters, which is in the observational limits of both VLTI and Keck. Moreover, in the observation of SgrA$^*$, the inclination angle is higher than $50\degree$.
In Figure \ref{fig:Constraint} (right), the fractional deviation parameter $(\delta)$ is plotted in the parameter space $a/M$ -- $p$ considering $\theta_{0}=50\degree$. The `BlueGreenYellow' region represents the points in the parameter space for which $\delta\geq -0.14$. Therefore, the parameter values corresponding to the white region are excluded from the observations in SgrA$^*$. Thus, for $p>0.24$, no values of $a$ can satisfy the bound on $\delta$. In summary, our rotating wormhole seems to be a viable alternative
since it tallies well with both shadow observations, for some range of 
values of the metric parameters.

\section{Conclusions}
 In this paper we first obtain a new rotating wormhole spacetime (Eq.~\ref{eq:rotating_wormhole_new}) starting from a static wormhole (Eq.~\ref{eq:static_wormhole_R0}) which has the features of a vanishing Ricci scalar ($R=0$) and the non-violation of the energy conditions in 
 a modified gravity scenario (a braneworld model). 
 Our rotating wormhole is the outcome of following the procedure of Azreg-A\"inou, as neither
the usual Newman-Janis algorithm nor its generalized version is successful in generating it. 
 
 \noindent The study of the geometry of the new rotating wormhole reveals that it falls in the class of Kerr-like wormholes rather than the Teo type wormholes. The  wormholes are characterized by three parameters which are the ADM mass M, spin parameter $a$, and the dimensionless parameter $p$. The throat radius also depends on these three parameters. The parameter $p$ signifies the deviation from Kerr black hole and carries the imprint of the braneworld gravity connection. In the limit $p\rightarrow 0$, the rotating wormhole turns out to be the Kerr black hole. However, in the limit $a\rightarrow 0$ the spacetime reduces to the $R=0$ static wormhole, though for the rotating version ($a\neq 0 $), $R\neq 0$. This is a striking feature. In fact, for a look-alike metric ansatz with the aim of obtaining $R=0$ rotating wormholes, one ends up finding the Kerr as the unique solution. Additionally, the smooth and continuous behaviour of the six Zakhary and McIntosh (ZM) invariants demonstrate the regular character of our metric. 

 \noindent Next, we have studied the effective energy-momentum tensor, resulting from the Einstein field equations i.e. $T_{\mu\nu}= \frac{1}{\kappa} G_{\mu\nu}$ (we choose units where $\kappa=1$), as the source {\color[rgb]{1,0,0} for} our rotating wormhole. 
 The diagonalization of such an effective energy-momentum tensor in a `locally non-rotating frame' (LNRF)
 gives the formulae for the effective energy density and anisotropic pressures which we use to study the energy conditions. In the framework of GR, our rotating wormhole violates all the energy conditions, as expected. The violations of energy conditions are graphically demonstrated in Figs. \ref{fig:WEC} and \ref{fig:NEC}. From the braneworld gravity perspective, it may however be possible to show, following \cite{skslssg}, that on-brane matter, as seen by an on-brane observer, does not violate the energy conditions, in the presence of an effective geometric stress energy attributed to the radion scalar field. However, to establish this for our rotating wormhole spacetime, we need to solve the scalar field equation assuming axial symmetry, which is likely to be a
 formidable task. We do hope to address this interesting issue
in future work.

\noindent Further, we study the null trajectories in the background of the new rotating wormhole. In particular, we obtain the shadow profiles. Fortunately, the related Hamilton-Jacobi equation for null geodesics is separable and we are able to apply the analytic treatment to obtain the shadow profiles. The shape and size of the shadow boundaries depend upon the parameters related to the wormhole geometry, which are $M$, $a$, and $p$, and also the external parameter, i.e. inclination angle $\theta_0$ between the observer line of sight and the rotation axis of the wormhole. Given the choice of $\theta_0$, the shape and size of the shadow boundaries in a plane $\alpha/M$ vs. $\beta/M$ depend only on two parameters: $a$ and $p$. Graphical demonstrations of such shadow profiles for various combinations of $a$ and $p$ are shown in Fig.~\ref{fig:Whsh}. A careful look at these figures reveal that the relative size of the shadow decreases as we increase $a$ or $p$ or both, but more prominently with an increase in $p$. The shape of the shadow boundaries deviates from the circle primarily with an increase in $a$ (although it depends upon $p$ and $\theta_0$ as well).  The shadow profiles in the bottom panel of Fig.~\ref{fig:Whsh} show that the deviation from Kerr shadows ($p=0$) is imprinted in the size of shadow profiles as we increase the wormhole parameter $p$ for any value of the rotation parameter. However, the deviation from the shape of Kerr shadows with a variation of $p$ is prominent only for higher values of $a$. Subsequently, we obtain restrictions on the metric parameters using EHT observations of M87$^*$ and SgrA$^*$ and the available bounds on deviation from circularity $(\Delta C)$ and fractional deviation parameter $(\delta)$. It is evident from Figure \ref{fig:Constraint} (left) that, for the shadow of M87$^*$, all theoretically allowed values of metric parameters respect the constraint on $\Delta C$ and thus the parameter space cannot be constrained. However, when the shadow data for SgrA$^*$ is used, the limit on $\delta$ bounds $p\leq 0.24$, which is also clear from Figure \ref{fig:Constraint} (right). 

\noindent Thus, the results from the analysis of shadow profiles throws up an interesting question--can the supermassive black holes  M87* or SgrA* be modeled as rotating wormholes (such as the one presented here) supported by a regular matter source in a modified gravity (braneworld) scenario? As mentioned above this is, at present,
mere speculation and requires more theoretical work. 
The current sensitivity of the EHT 
and the available data seem to be inadequate to remove the existing degeneracy among plausible candidates (such as Kerr or other geometries like ours here) 
which, as of now, may all be good enough in serving as models for such compact objects. Therefore, better sensitivity and more data are certainly required in order to pinpoint the right geometry for a given compact object. 
From an exclusively theoretical standpoint, our work here may have
two implications, in future. Firstly, we have found a regular, rotating, asymptotically flat wormhole spacetime which, at large distances from the throat, approaches Kerr geometry. This, by itself, is new since it addresses and resolves issues which arose earlier while constructing rotating versions 
of the given non-rotating spacetime. Secondly, if we are able to prove in future that our solution works (with viable matter) in a braneworld gravity (or any other alternative gravity) context, it may provide
useful motivation and additional support for pursuing such a modified theory of gravity.

\section*{Acknowledgements}
\noindent AK thanks Rohan Pramanick for some useful discussions. The research of SJ is partially funded by the SERB, DST, Government of India, via TARE fellowship award no. TAR/2021/000354, administered by the Department of Physics, Indian Institute of Technology Kharagpur, India.
The research of AK is supported through a fellowship by the Indian Institute
of Technology Kharagpur, India.

\end{document}